\documentclass[traditabstract]{aa}

\usepackage{graphicx}
\usepackage{caption}
\usepackage{txfonts}
\usepackage{natbib}
\usepackage{enumerate}
\usepackage{bm}
\usepackage{hyperref}
\hypersetup{
  colorlinks=true,
  linkcolor=blue,
  citecolor=blue,
  urlcolor=blue,
}

\bibpunct{(}{)}{;}{a}{}{,} 

\newcommand{\sfr}{M$_\odot$\,yr$^{-1}$}
\newcommand{\rhostar}{$\rho_*$}

\newcommand{\kms}{km\,s$^{-1}$}

\newcommand{\htwo}{H$_2$}

\newcommand{\lir}{L$_{\rm IR}$}
\newcommand{\lnu}{L$_{\nu}$}
\newcommand{\av}{A$_V$}
\newcommand{\grasil}{{\sc grasil}}
\newcommand{\spit}{{\it Spitzer}}
\newcommand{\hers}{{\it Herschel}}

\newcommand{\mstar}{$M_{\rm star}$}
\newcommand{\zmed}{$z_{\rm med}$}
\newcommand{\mlow}{$M_{\rm low}$}

\newcommand{\mburst}{$M_{\rm burst}$}
\newcommand{\mdust}{$M_{\rm dust}$}
\newcommand{\mgas}{$M_{\rm gas}$}
\newcommand{\tdust}{$T_{\rm dust}$}
\newcommand{\hi}{\rm H{\sc i}}

\newcommand{\ha}{${\rm H}\alpha$}

\newcommand{\lya}{${\rm Ly}\alpha$}
\newcommand{\micron}{$\mu$m}

\def\tex {\ifmmode{{T}_{\rm ex}}\else{$T_{\rm ex}$}\fi}
\def\tmb {\ifmmode{{T}_{\rm mb}}\else{$T_{\rm mb}$}\fi}
\def\ci  {\ifmmode{{\rm C}{\rm \small I}}\else{C\ts {\scriptsize I}}\fi}
\def\hi  {\ifmmode{{\rm H}{\rm \small I}}\else{H\ts {\scriptsize I}}\fi}
\def\hh  {\ifmmode{{\rm H}_2}\else{H$_2$}\fi}
\def\kms    {\ifmmode{{\rm \ts km\ts s}^{-1}}\else{\ts km\ts s$^{-1}$}\fi}
\def\msun   {\ifmmode{{\rm M}_{\odot}}\else{M$_{\odot}$}\fi}
\def\msunpc   {\ifmmode{{\rm M}_{\odot}\,{\rm pc}^{-2}}\else{M$_{\odot}$\,pc$^{-2}$}\fi}
\def\msunyr   {\ifmmode{{\rm M}_{\odot}\,{\rm yr}^{-1}}\else{M$_{\odot}$\,yr$^{-1}$}\fi}
\def\lsun   {\ifmmode{{\rm L}_{\odot}}\else{L$_{\odot}$}\fi}
\def\zsun   {\ifmmode{{\rm Z}_{\odot}}\else{Z$_{\odot}$}\fi}

\begin{document}

\title{New light on gamma-ray burst host galaxies with Herschel
\thanks{{\it Herschel} is an ESA space observatory with science instruments
provided by European-led Principal Investigator consortia and with important
participation from NASA.}} 

\author{L.~K. Hunt \inst{\ref{inst:hunt}}
\and
E.~Palazzi\inst{\ref{inst:palazzi}}
\and
M.~J.~Micha{\l}owski\inst{\ref{inst:gent},\ref{inst:roe}}\thanks{FWO Pegasus Marie Curie Fellow}
\and
A.~Rossi\inst{\ref{inst:palazzi},\ref{inst:rossi}}
S.~Savaglio\inst{\ref{inst:savaglio1}} 
\and
S.~Basa\inst{\ref{inst:basa}}  
\and
S.~Berta\inst{\ref{inst:savaglio1}}
\and
S.~Bianchi\inst{\ref{inst:hunt}}
\and
S.~Covino\inst{\ref{inst:covino}}
\and
V.~D'Elia\inst{\ref{inst:elia1},\ref{inst:elia2}} 
\and
P.~Ferrero\inst{\ref{inst:fer}}
\and
D.~G\"otz\inst{\ref{inst:saclay}}
\and
J.~Greiner\inst{\ref{inst:savaglio1}} 
\and
S.~Klose\inst{\ref{inst:rossi}}
\and
D.~Le Borgne\inst{\ref{inst:leb}} 
\and
E.~Le Floc'h\inst{\ref{inst:saclay}} 
\and
E.~Pian\inst{\ref{inst:palazzi}}
\and
S.~Piranomonte\inst{\ref{inst:elia2}} 
\and
P.~Schady\inst{\ref{inst:savaglio1}} 
\and
S.~D.~Vergani\inst{\ref{inst:vergani},\ref{inst:covino}}
}

\offprints{L. K. Hunt}
\institute{INAF - Osservatorio Astrofisico di Arcetri, Largo E. Fermi, 5, 50125, Firenze, Italy
\label{inst:hunt}
\email{hunt@arcetri.astro.it} 
\and
INAF-IASF Bologna, Via Gobetti 101, I-40129 Bologna, Italy \label{inst:palazzi}
\and
Sterrenkundig Observatorium, Universiteit Gent, Krijgslaan 281-S9, 9000, Gent, Belgium  \label{inst:gent}
\and
SUPA\thanks{Scottish Universities Physics Alliance}, Institute for Astronomy, 
University of Edinburgh, Royal Observatory, Edinburgh, EH9 3HJ, UK 
\label{inst:roe}
\and
Th\"uringer Landessternwarte Tautenburg, Sternwarte 5, D-07778 Tautenburg, Germany \label{inst:rossi}
\and
Max-Planck-Institut f\"{u}r Extraterrestrische Physik, Giessenbachstra{\ss}e, D-85748 Garching bei M\"{u}nchen, Germany \label{inst:savaglio1}
\and
Aix Marseille Universit\'e, CNRS, LAM (Laboratoire d'Astrophysique de Marseille) UMR 7326, 13388, Marseille, France \label{inst:basa}
\and
INAF/Osservatorio Astronomico di Brera, via Emilio Bianchi 46, 23807 Merate (LC), Italy \label{inst:covino}
\and
ASI Science Data Centre, Via Galileo Galilei, 00044 Frascati (RM), Italy \label{inst:elia1}
\and
INAF - Osservatorio Astronomico di Roma, Via di Frascati, 33, 00040 Monteporzio Catone, Italy \label{inst:elia2}
\and
Instituto de Astrofísica de Andalucía (IAA-CSIC), Glorieta de la Astronomía s/n, E-1008, Granada, Spain \label{inst:fer}
\and
Laboratoire AIM-Paris-Saclay, CEA/DSM/Irfu - CNRS - Universit\'e Paris Diderot, CE-Saclay, pt courrier 131, F-91191 Gif-sur-Yvette, France \label{inst:saclay}
\and
Institut d'Astrophysique de Paris, UMR 7095, CNRS, UPMC Univ. Paris 06, 98bis boulevard Arago, F-75014 Paris, France \label{inst:leb}
\and
GEPI-Observatoire de Paris, CNRS UMR 8111, Univ. Paris-Diderot, 5 Place Jules Jannsen, F-92190 Meudon, France \label{inst:vergani}
}

\date{Received  2013/ Accepted  2013}

\titlerunning{Dust in GRB host galaxies}
\authorrunning{Hunt et al.}

\abstract{Until recently, dust emission has been detected in very few
host galaxies of gamma-ray bursts (GRBHs). With \hers, we have now observed
17 GRBHs up to redshift $z\sim3$ and detected seven of them at infrared (IR) wavelengths. 
This relatively high detection rate (41\%) may be due to
the composition of our sample which at a median redshift of 1.1
is dominated by the hosts of dark GRBs.
Although the numbers are small, statistics suggest that dark GRBs are more likely
to be detected in the IR than their optically-bright counterparts.
Combining our IR data with optical, near-infrared, and radio data from 
our own datasets and from the literature,
we have constructed spectral energy distributions (SEDs) which span up to 6 orders
of magnitude in wavelength.
By fitting the SEDs, we have obtained stellar masses,
dust masses, star-formation rate (SFR), and extinctions for our sample galaxies. 
We find that GRBHs are galaxies that tend to have a high specfic SFR (sSFR), and
like other star-forming galaxies,
their ratios of dust-to-stellar mass are well correlated with sSFR.
Dust masses of GRBHs relative to stellar mass and SFR fall within the range
of other star-forming galaxies in the local universe, and of
sub-millimeter galaxies (SMGs) and luminous IR galaxies for redshift $z\ga1$.
We incorporate our \hers\ sample into a larger compilation of GRBHs, after
checking for consistency in mass and SFR estimations.
This combined sample is compared to SFR-weighted median stellar masses of the widest,
deepest galaxy survey to date in order to establish whether or not GRBs can
be used as an unbiased tracer of cosmic comoving SFR density (SFRD) in the universe.
In contrast with previous results, this comparison shows that GRBHs are medium-sized
galaxies with relatively high sSFRs, as might be expected for galaxies selected
on the basis of SFR because of the explosive GRB event.
Stellar masses and sSFRs of GRBHs as a function of redshift are similar to what is expected for 
star-forming galaxy populations at similar redshifts.
We conclude that there is no strong evidence that GRBs are biased tracers of SFRD;
thus they should be able to reliably probe the SFRD to early epochs.
}
\maketitle


\section{Introduction}
\label{sec:intro}
Long-duration gamma-ray bursts (GRBs) 
are so luminous that they can shine through highly
obscured galaxies \citep[e.g.,][]{djor01}
and can be seen even at very high redshifts 
\citep{salvaterra09,tanvir09,cucchiara11}.
They are thought to originate in the collapse of very massive stars
at the end of their evolution \citep{pacz98,macfayden99,woosley06}.
Because of this association with massive stars, GRBs have recently been used, 
thanks to the advent of the dedicated mission {\it Swift}, to infer the cosmic evolution 
of the star formation rate density (SFRD) up to $z \sim 9$ 
\citep{yuksel08,kistler09,butler10,robertson12,elliott12,trenti13}.

Although GRBs are rare events, they
enable identification of galaxies 
that would not otherwise be singled out even in deep flux-limited surveys,
thus making GRBs a potentially powerful probe of galaxy evolution.
Galaxies hosting GRBs (GRBHs) are better known in the low-$z$ regime ($z\la 1.5$), 
where they are typically low-mass, young, star-forming
blue galaxies with low dust extinction \citep{lefloch03,christensen04,fruchter06,savaglio09}.
However, these galaxy characteristics, possibly related to a
selection bias because of the consideration of only optically bright GRB afterglows,
may not be so uniform at high redshift. 
Indeed,
evidence is mounting that the GRBH population is much more diverse 
at $z\ga 1.5$ than previously thought.

Dark GRBs,
those for which the observed optical afterglow is very faint
relative to the extrapolation from the X-ray 
\citep{jakobsson04,vanderhorst09},
tend to be found in massive, star-forming galaxies
with red colors, high extinction and large SFRs 
\citep[e.g.,][]{kruhler11,rossi12,perley13}.
Dark GRBs comprise 
up to 30-40\% of the {\it Swift} GRB dataset \citep{fynbo09,greiner11,melandri12},
and thus the assumption that all GRBHs are low-mass, metal poor galaxies 
may be an oversimplification. 
Consequently, theoretical work based on 
this assumption using GRBHs as cosmological probes \citep{campisi09,niino11,robertson12}
could be undermined. 

Until recently, statistics on dark GRBs and their host galaxies have been poor
\citep[e.g.,][]{kann10}.
Because of their optical faintness, it has been very difficult to localize
the optical afterglow of dark GRBs, and thus identify the host 
\citep[e.g.,][]{rossi12}.
Now, thanks to sustained observational efforts, 
we have a considerably better understanding of dark GRBs and their host galaxies.
In most dark GRBs, the optical faintness is caused
by high dust extinction columns and moderate redshift 
\citep{perley09,greiner11,kruhler11,rossi12,melandri12,covino13}.
However, the properties of dust extinction are not the same for all GRBs, although
there is no clear evidence that 
afterglow extinction curves differ significantly from those commonly used 
\citep{perley08,zafar11,zafar12,schady12}.
It is thus difficult to make conclusive statements about the nature
of GRB hosts \citep[e.g.,][]{delia13,elliott13};
it is neither true that all massive, metal-rich GRBHs are found from dark GRBs
\citep[e.g.,][]{lefloch02}, nor do
all dark GRBs reside in massive hosts at high redshift. 

%
\begin{center}
\begin{table*}
      \caption[]{Host galaxy sample} 
\label{tab:sample}
\resizebox{\linewidth}{!}{
{
\tiny
\begin{tabular}{llllccll}
\hline
\multicolumn{1}{c}{GRB} & 
\multicolumn{1}{c}{RA} & 
\multicolumn{1}{c}{Dec.} & 
\multicolumn{1}{l}{Redshift} & 
\multicolumn{1}{c}{Redshift} & 
\multicolumn{1}{c}{Dark?$^{\mathrm a}$} &
\multicolumn{2}{c}{Observing ID} \\ 
& \multicolumn{2}{c}{(J2000)$^{\mathrm{b}}$} &
&
\multicolumn{1}{l}{Reference\,$^{\mathrm{c}}$} && \multicolumn{1}{c}{PACS} & \multicolumn{1}{c}{SPIRE} \\ 
\hline
\\
970828                & 18:08:31.56 & +59:18:51.2   &  0.958  & 1 & Yes &  1342258609,10 & 1342241152 \\ 
980613                & 10:17:57.9  & +71:27:26.3   &  1.097  & 2 & No  &  1342270864,65 & \multicolumn{1}{c}{$-$} \\ 
980703                & 23:59:06.67 & +08:35:06.7   &  0.966  & 3 & No  &  \multicolumn{1}{c}{$-$}           & 1342212419 \\ 
990705                & 05:09:54.50 & $-$72:07:53.0   &  0.842  & 4 & No  &  1342261814,15 & 1342214749 \\ 
020127                & 08:15:01.42 & +36:46:33.9   &  1.9$^{\mathrm d}$    & 5 & Yes &  1342243296,97 & 1342251906 \\ 
020819B               & 23:27:19.52 & +06:15:53.2   &  0.411  & 6 & Yes &  1342246708,09 & 1342212294 \\ 
030115                & 11:18:32.63 & +15:02:59.9   &   2.0$^{\mathrm e}$   & 7  & Yes & 1342247632,33 & \multicolumn{1}{c}{$-$} \\ 
050219A               & 11:05:39.07 & $-$40:41:04.6   &  0.2115  & 8 & Yes  & 1342248286,87 & 1342247261 \\ 
050223                & 18:05:32.99 & $-$62:28:18.8   &  0.584  & 9  & Yes & 1342243788,89 & 1342214756 \\ 
051022                & 23:56:04.1  & +19:36:24.1   &  0.807  & 6 & Yes  & 1342238085,86 & 1342247985 \\ 
060904A               & 15:50:54.56 & +44:59:10.5   &  2.55$^{\mathrm d}$  & 11 & Yes & 1342261849,50 & \multicolumn{1}{c}{$-$} \\ 
070306                & 09:52:23.3  & +10:28:55.5   &  1.496  & 12 & Yes & 1342254142,43$^{\mathrm{f}}$ & \multicolumn{1}{c}{$-$} \\ 
071021                & 22:42:34.31 & +23:43:06.5   &  2.452  & 13 & Yes & 1342246178,79 & 1342258356 \\ 
080207                & 13:50:03.01 & +07:30:07.8   &  2.086  & 13 & Yes & 1342257547,48 & 1342261526 \\ 
080325                & 18:31:34.3  & +36:31:24.8   &  1.78   & 14 & Yes & 1342245672,73 & 1342241156 \\ 
090404                & 15:56:57.52 & +35:30:57.5   &  3.00$^{\mathrm d}$  & 14 & Yes & 1342258433,34 & 1342241163 \\ 
090417B               & 13:58:46.66 & +47:01:04.4   &  0.345  & 15 & Yes & 1342257593,94 & 1342259467 \\ 
\\
\hline
\end{tabular}
}
}
\vspace{0.5\baselineskip}
\begin{description}
\item
[$^{\mathrm{a}}$]
These GRBs are classified as dark according to the definitions by
\citet[][]{jakobsson04,vanderhorst09}.
\item
[$^{\mathrm{b}}$]
These are the positions of the host galaxy, which may or may not be
exactly coincident with the position of the GRB afterglow.
\item
[$^{\mathrm{c}}$] 
(1) \citet{djor01};
(2) \citet{djor03};
(3) \citet{djor98};
(4) \citet{lefloch02};
(5) \citet{berger07};
(6) \citet{levesque10b};
(7) \citet{levan06};
(8) \citet{rossi14};
(9) \citet{pellizza06};
(10) \citet{chary07};
(11) \citet{xiao11};
(12) \citet{jaunsen08};
(13) \citet{kruhler12};
(14) \citet{perley13};
(15) \citet{holland10}.
\item
[$^{\mathrm{d}}$]
Photometric redshift.
\item
[$^{\mathrm{e}}$]
Photometric redshift determined by our \grasil\ fits.
\item
[$^{\mathrm{f}}$]
Although this source was in our target list,
this observation was acquired in the OT2 proposal,
OT2\_ppschady\_2.
\end{description}
\end{table*}
\end{center}

Direct observations of the entire spectral energy distribution
(SED) of dark GRBHs would help to place them
in the context of other high-$z$ galaxy populations, 
especially since 
the most common understanding of the high-$z$ universe 
is based on optically-selected (rest-frame UV) galaxy surveys.
So far, only a handful of GRBHs have been detected with sub-millimeter (submm) facilities 
\citep{barnard03,berger03,tanvir04,priddey06,wang12,michalowski13}.
However, in this minority 
SFRs can be high, $\sim$500\,\sfr, as high as those of the submm galaxy (SMG) population
\citep{chapman04,greve05}
even though the
galaxies picked out by the two selection criteria are quite different. 
There has been no CO emission found in any of the GRBHs observed so far
\citep[e.g.,][]{hatsukade11} 
and there is some hint that dust in GRBHs
may be warmer than in typical ultra-luminous infra-red galaxies (ULIRGs) and 
SMGs \citep[][]{priddey06,michalowski08}.

To better understand the properties of dark GRB hosts,
and to assess their potential impact on the GRBH population in general,
we undertook an observational campaign with \hers;
this paper presents results from this campaign.
We characterize the dust content, stellar mass, and SFRs of GRBHs through
\hers\ observations of 17 GRBHs, 14 of which host dark GRBs.
This is the first time that \hers\ has been used to examine
dust in GRBHs; dust emission is detected in seven of our targets. 
Sample selection is described in Sect. \ref{sec:sample}, and
Sect. \ref{sec:data} reports the \hers\ observations and the other data
incorporated in the compilation of the SEDs, together with the 
procedures for the photometry.
SED fitting is discussed in Sect. \ref{sec:grasil}, and 
Sect. \ref{sec:results} gives the results of the fitting 
in terms of stellar masses, dust masses, SFRs, and dust extinction.
The properties of our sample of GRBHs are compared with other
GRBH samples and other high-$z$ star-forming galaxy populations in Sect. \ref{sec:context}.
Throughout the paper, we assume a 
$\Omega_m$\,=\,0.3, $\Omega_\Lambda$\,=\,0.7 cosmology, with 
Hubble constant $H_0$\,=\,70\,km\,s$^{-1}$\,Mpc$^{-1}$.

\section{Sample selection}
\label{sec:sample}

The \hers\ target list is based on 118 GRBHs imaged
with \spit/IRAC \citep[InfraRed Array Camera,][]{fazio04}, available in June, 2010.
We retrieved these images from the \spit\ archive, and
when available, we also retrieved images at 24\,\micron\ acquired with
\spit/MIPS \citep[Multiband Imaging Photometer,][]{rieke04}.
Both sets of images were reduced with MOPEX \citep{mopex}, taking
into account the difference between the older IRAC images and those acquired
with warm \spit.
After having performed photometry on the images for
this parent sample, we defined the \hers\ observing sample by requiring that
the host galaxy be detected in at least two \spit\ bands (usually IRAC).
To ensure detection with \hers, based on normal galaxy SEDs we estimated that
the IRAC 3.6\,\micron\ or 4.5\,\micron\ flux needed to be $\ga$10\,$\mu$Jy, so selected only
galaxies that fulfilled this flux limit.
Finally, we avoided targets in crowded fields, so that the \hers\ photometry
would not be contaminated by extraneous objects near the hosts.
 
We thus obtained a sample of 17 GRBHs, which were observed over both (OT1 and OT2)
\hers\ observing cycles.
An additional host observed in OT1, GRB\,980425, 
the closest GRBH at $z\,=\,0.0085$, is discussed by \citet{michalowski13}.
In OT1, we included both optically-bright and dark GRBs,
while in OT2, the targets were required to be the hosts of dark GRBs;
dark bursts thus comprise the bulk of our sample (14 of 17 targets host dark GRBs).
Throughout the paper, we define a dark GRB as one that optically falls short
of the prediction of the fireball model, namely with an optical-to-X-ray spectral
index $\beta_{\rm ox}<$0.5 \citep{jakobsson04}.
Table \ref{tab:sample} gives the host-galaxy positions and redshifts of the \hers\ targets.
Redshifts range from $z\,=\,0.21$ (GRB\,050219A) to 
$z\,=\,3$ (GRB\,090404); the median redshift \zmed\,=\,1.1.
The uncertainties in the GRBH positions are $\la$0\farcs5 in all cases.

\section{The data and the photometry}
\label{sec:data}

We have acquired \hers\ \citep{pilbratt10}
PACS and SPIRE maps for the 17 GRBHs in our observing sample,
and combined them with our own data and with data from the literature to compile
SEDs for GRB host galaxies that span almost 3 orders of magnitude in wavelength.

\subsection{{\it Herschel} observations}
\label{sec:herschel}

Through two open-time observing programs (OT1\_lhunt\_2 and OT2\_lhunt\_3)
we obtained images at 100 and 160\,\micron\ with 
PACS \citep[Photodetector Array Camera 
\& Spectrometer\footnote{PACS has been developed by a consortium of institutes led by MPE
(Germany) and including UVIE (Austria); KU Leuven, CSL, IMEC (Belgium); CEA,
LAM (France); MPIA (Germany); INAF-IFSI/OAA/OAP/OAT, LENS, SISSA (Italy); IAC
(Spain). This development has been supported by the funding agencies BMVIT
(Austria), ESA-PRODEX (Belgium), CEA/CNES (France), DLR (Germany), ASI/INAF
(Italy), and CICYT/MCYT (Spain).},][]{poglitsch10} 
and at 250, 350, and 500\,\micron\ with
SPIRE \citep[Spectral and Photometric Imaging REceiver\footnote{SPIRE has 
been developed by a consortium of institutes led by Cardiff
University (UK) and including Univ. Lethbridge (Canada); NAOC (China); CEA, LAM
(France); IFSI, Univ. Padua (Italy); IAC (Spain); Stockholm Observatory
(Sweden); Imperial College London, RAL, UCL-MSSL, UKATC, Univ. Sussex (UK); and
Caltech, JPL, NHSC, Univ. Colorado (USA). This development has been supported
by national funding agencies: CSA (Canada); NAOC (China); CEA, CNES, CNRS
(France); ASI (Italy); MCINN (Spain); SNSB (Sweden); STFC, UKSA (UK); and NASA
(USA).},][]{griffin10}.

We used PACS in Small-Scan map mode (20\arcsec/s),
with 10 scan legs, 3\arcmin\ long, separated by 4\arcsec\ steps.
The scans were divided into two  Astronomical Observation Requests (AORs), 
with orthogonal scan directions which were executed sequentially (see Table \ref{tab:sample}). 
With this configuration we obtain homogeneous coverage over an area with 
a diameter of $\sim2$\arcmin, sufficient to cover the region subtended
by the hosts.
Cross scans gave the needed redundancy to 
avoid 1/f noise and spurious detector glitches 
on science and noise maps.
The estimated 1$\sigma$ sensitivity is 0.5\,mJy at 100\,\micron\
and 1.7\,mJy at 160\,\micron.
With SPIRE in Small-Map Mode, we used 4 repetitions in order to obtain
a sensitivity of roughly the 1$\sigma$ confusion limit
\citep[see][]{nguyen10} of $\sim$6\,mJy beam$^{-1}$ at 250\,$\mu$m.
SPIRE observations were obtained for only a subset of the observations.

Data reduction for PACS and SPIRE was performed with {\sc Hipe} \citep[Herschel
Imaging Processing Environment;][]{hipe} v10.0. 
For PACS, the ``deep survey point-source" option was used, with masking performed on the images
themselves before combining the repetitions and orthogonal scans into a single map. 
We used pixel sizes of 2\farcs0 and 3\farcs0 for PACS 100, 160\,$\mu$m,
and 4\farcs5, 6\farcs25 and 9\farcs0 for SPIRE 250, 350, and 500\,$\mu$m, respectively.
These pixels well sample
the PACS and SPIRE full-width half-maximum beam sizes of 
$\sim$6\farcs8 and 11\farcs4 (for PACS 100 and 160\,\micron), and
18\farcs2, 24\farcs9, and 36\farcs3 (for
SPIRE 250, 350, and 500\,\micron, respectively). 
With the aim of maximizing sensitivity to extended emission,
we also reduced the PACS data with {\tt scanamorphos}
\citep{roussel13}.
However, the slightly larger reconstructed beam resulted in overall worse
noise characteristics although flux levels did not differ significantly
with the previous reduction.
Hence, we used the photometry from the deep-survey reduction mode.

\begin{figure*}[ht]
\centerline{
\includegraphics[angle=0,height=0.25\linewidth]{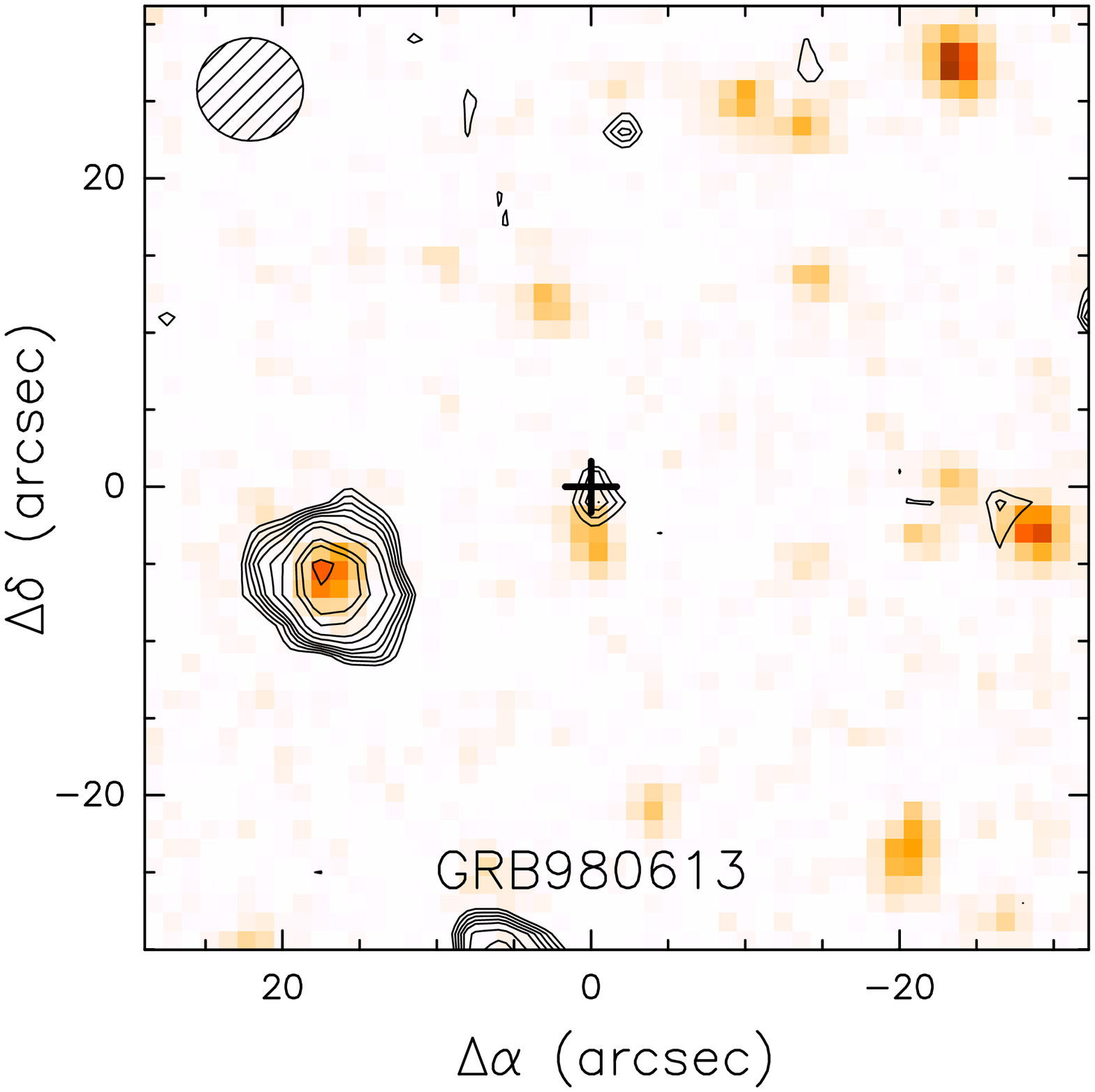}
\includegraphics[angle=0,height=0.25\linewidth]{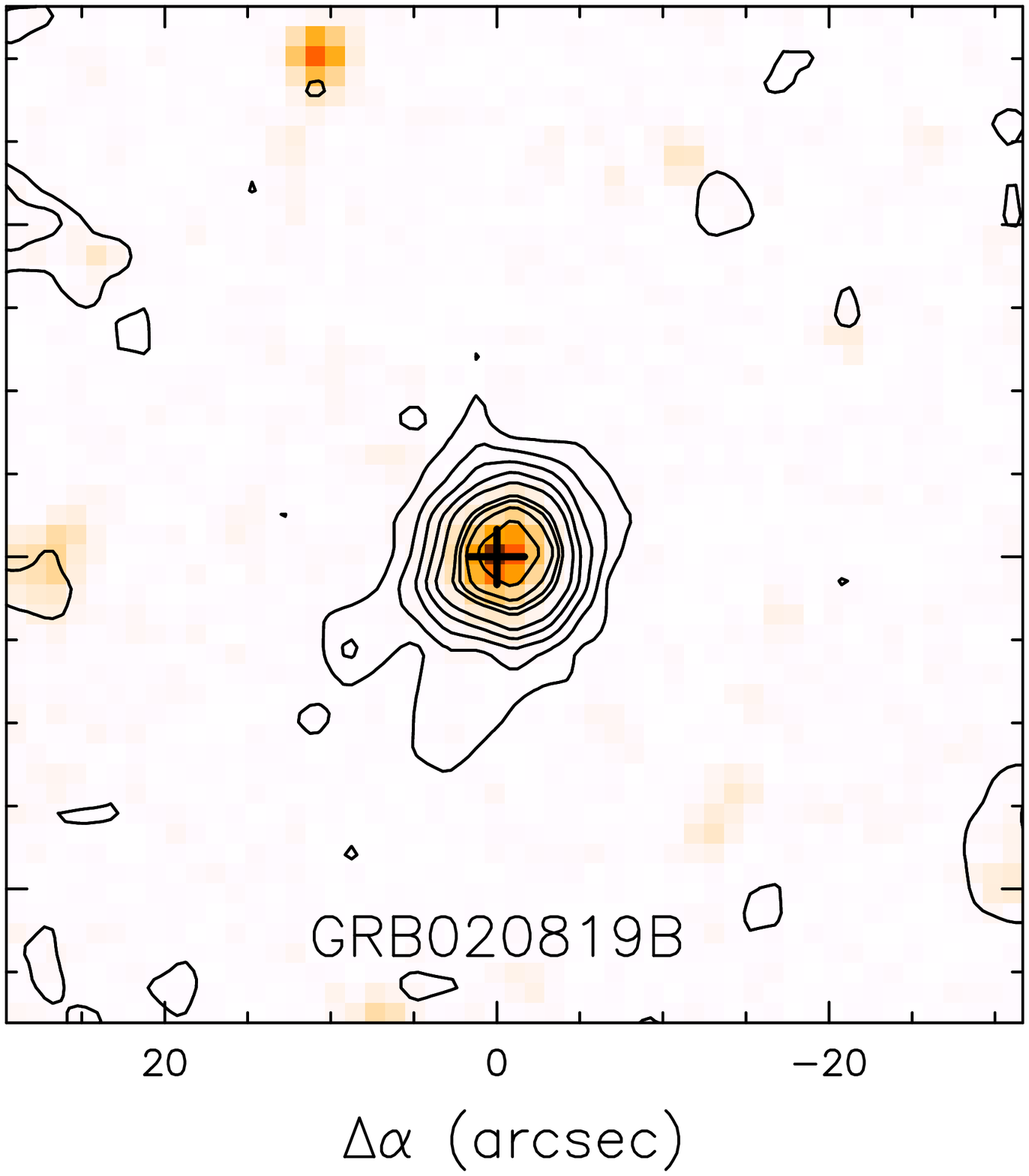}
\includegraphics[angle=0,height=0.25\linewidth]{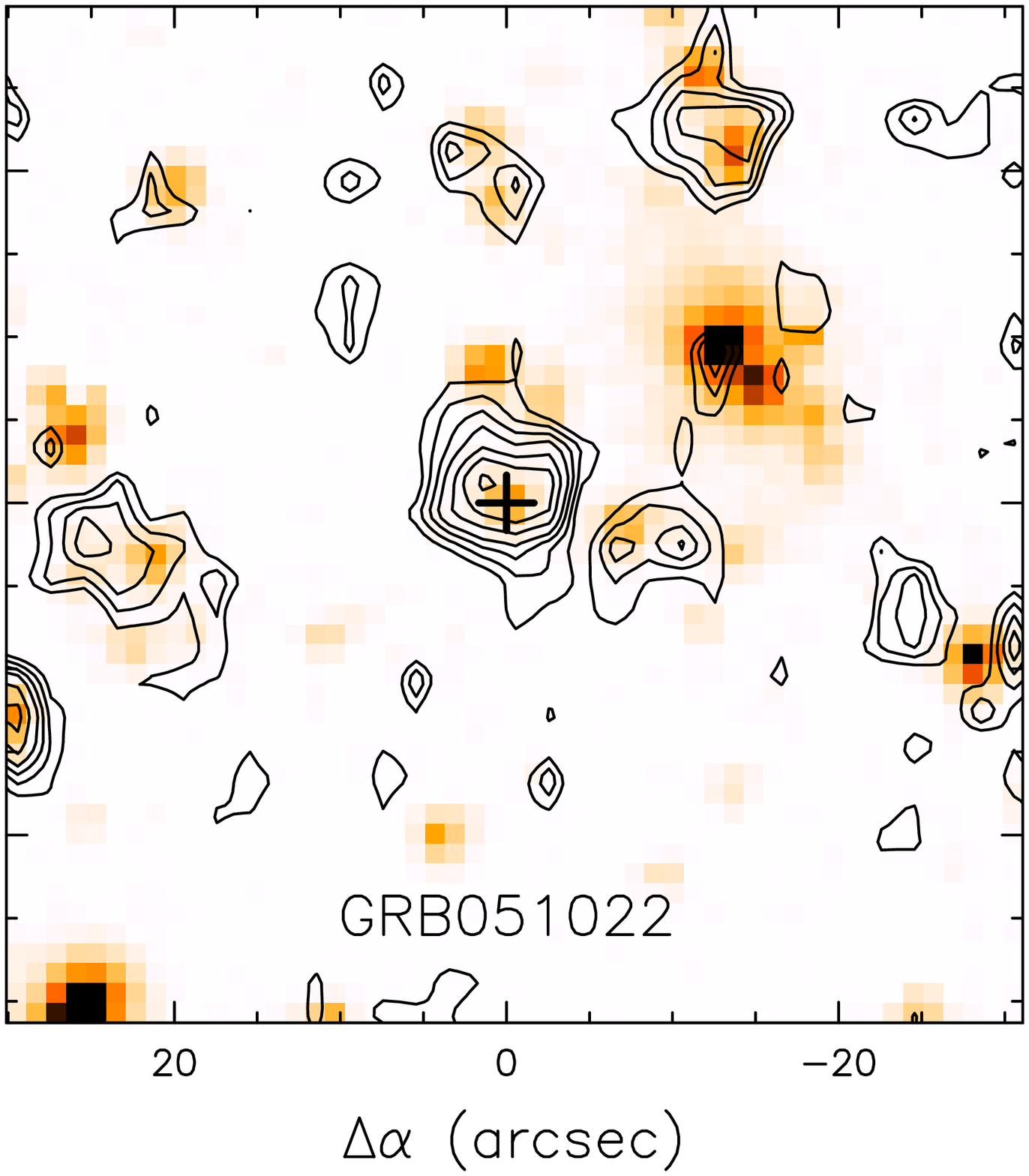}
\includegraphics[angle=0,height=0.25\linewidth]{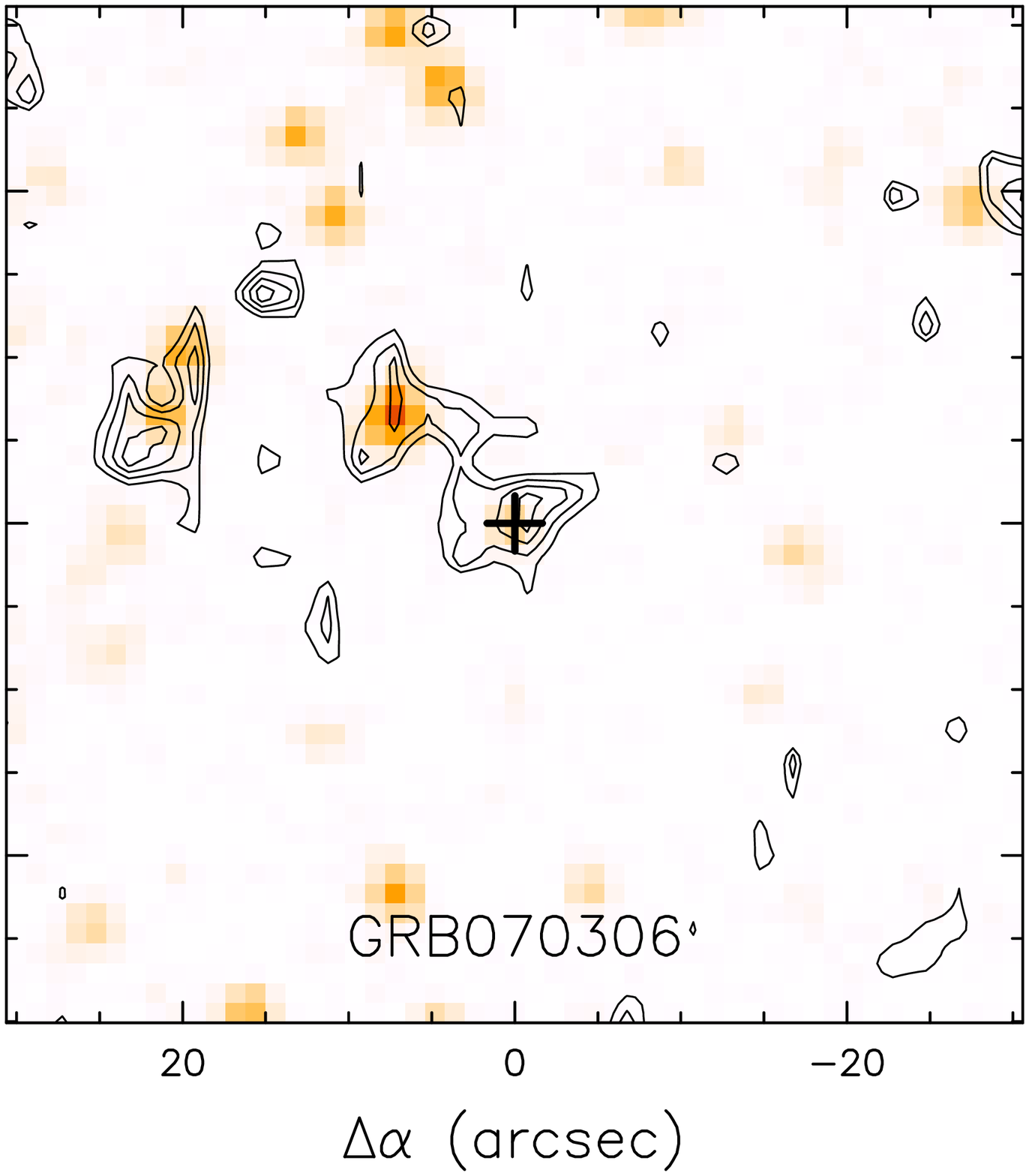}
}
\vspace{\baselineskip}
\centerline{
\includegraphics[angle=0,height=0.25\linewidth]{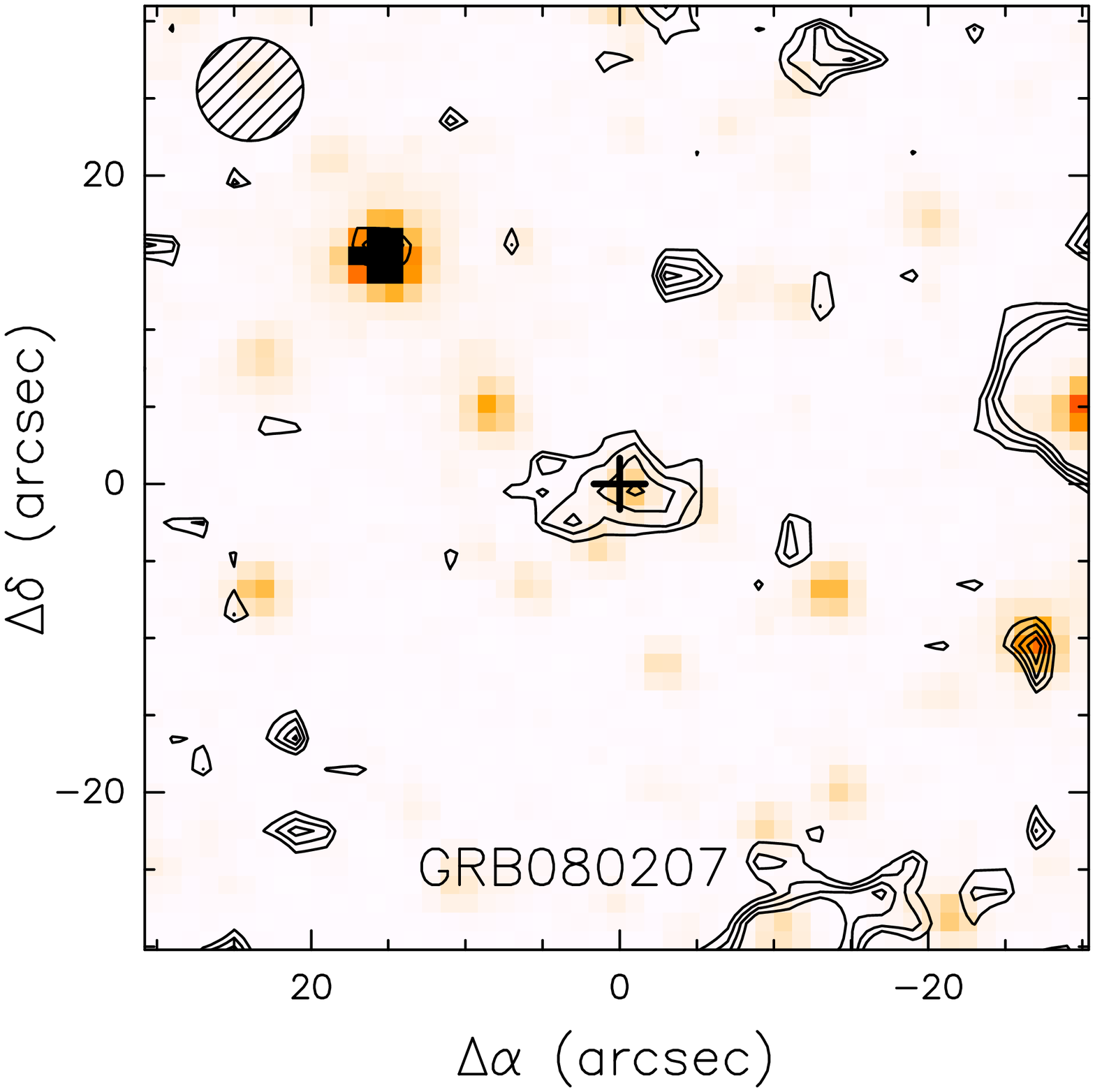}
\includegraphics[angle=0,height=0.25\linewidth]{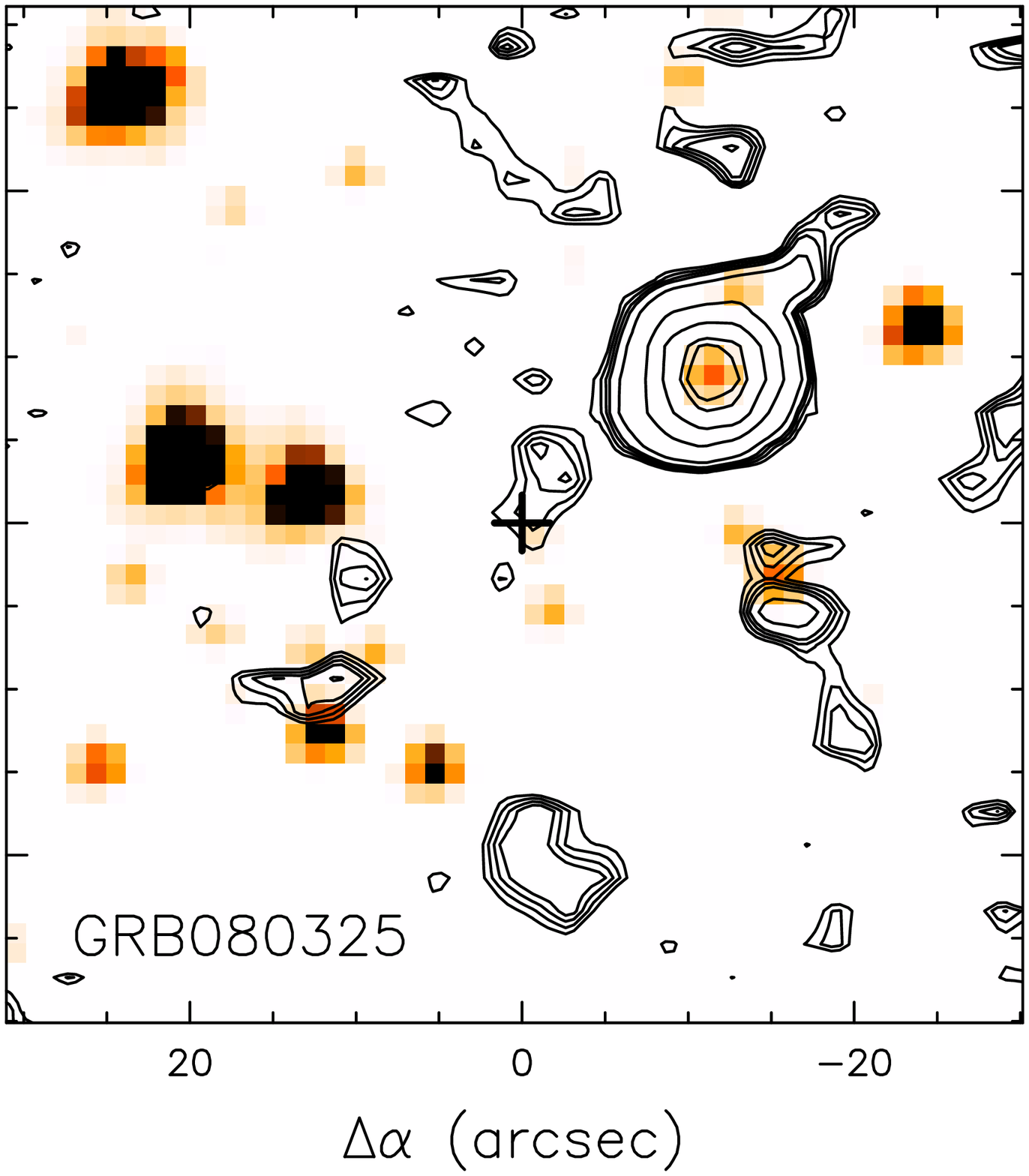}
\includegraphics[angle=0,height=0.25\linewidth]{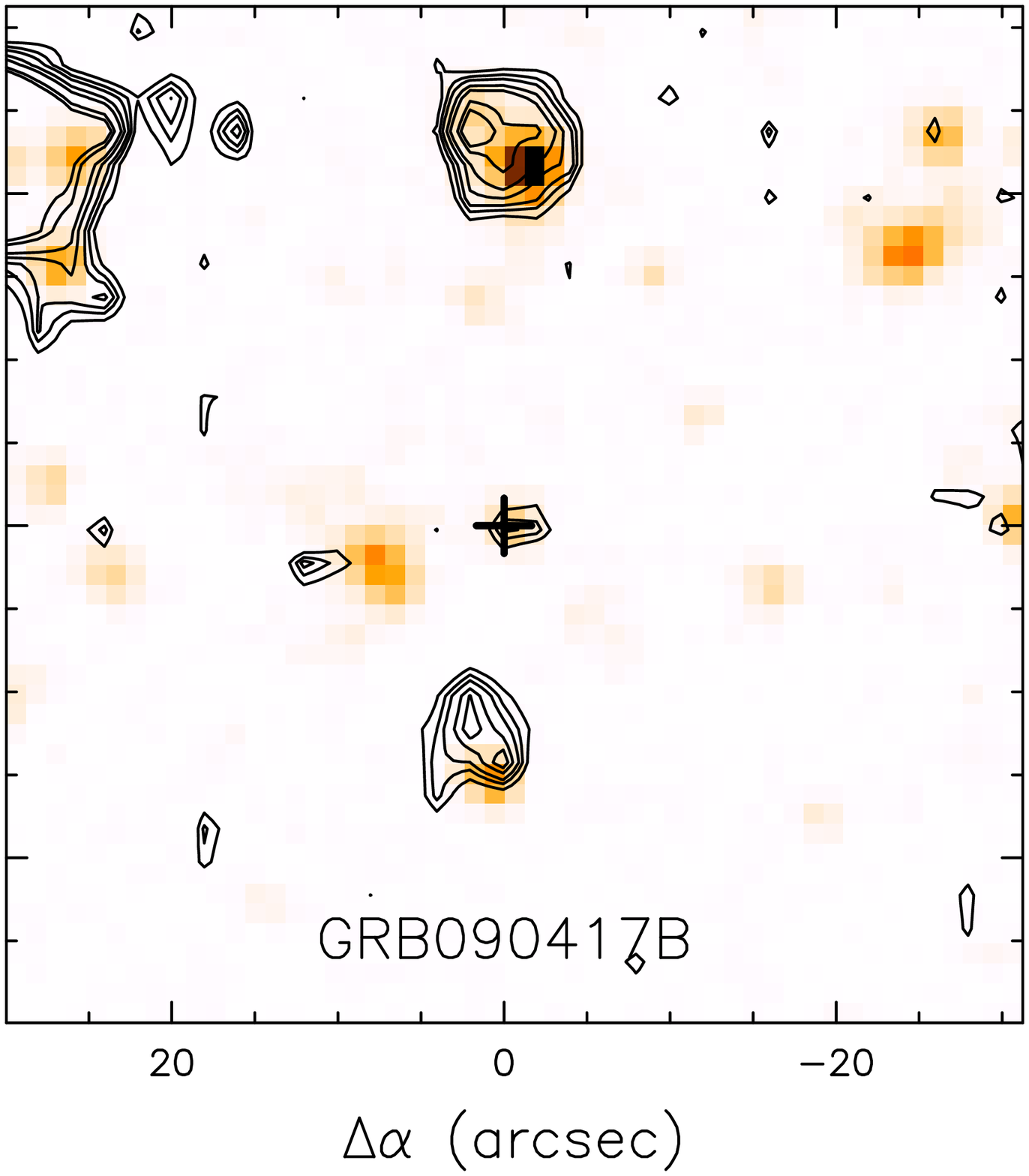}
}
\caption{PACS 100\,\micron\ images of the 7 detected galaxies superimposed as contours
on the \spit/IRAC 3.6\,\micron\ images shown with false colors (for the host of GRB\,980613,
the IRAC 4.5\,\micron\ is plotted).
The (0,0) position corresponds to the coordinates given in Table \ref{tab:sample}.
The slight vertical offsets in the various panels result from inability to perfectly center
the IRAC images because of the integer pixels.
The PACS 100\,\micron\ beam size is shown in the upper left corner of the left-most panel
in each row.
PACS contours start at 3$\sigma$ and run to 
4.5$\sigma$ (980613),
30$\sigma$ (020819B),
7$\sigma$ (051022),
6$\sigma$ (070306),
6$\sigma$ (080207),
3.5$\sigma$ (080325), and
5$\sigma$ (090417B).
These $\sigma$ values per pixel correspond to the correlated noise measured from the images, 
which are 
roughly 4 times smaller than the true noise (see PACS documentation
and text).
A $+$ marks the nominal GRBH position (see Table \ref{tab:sample}).
The detection of the host of GRB\,080325 is only marginal, 
significant at $\sim 2.5\sigma$.
For display purposes, the PACS images have been rebinned to
smaller pixel sizes.
}
\label{fig:herschel}
\end{figure*}

\begin{figure*}[ht!]
\centerline{
\includegraphics[angle=0,width=13cm]{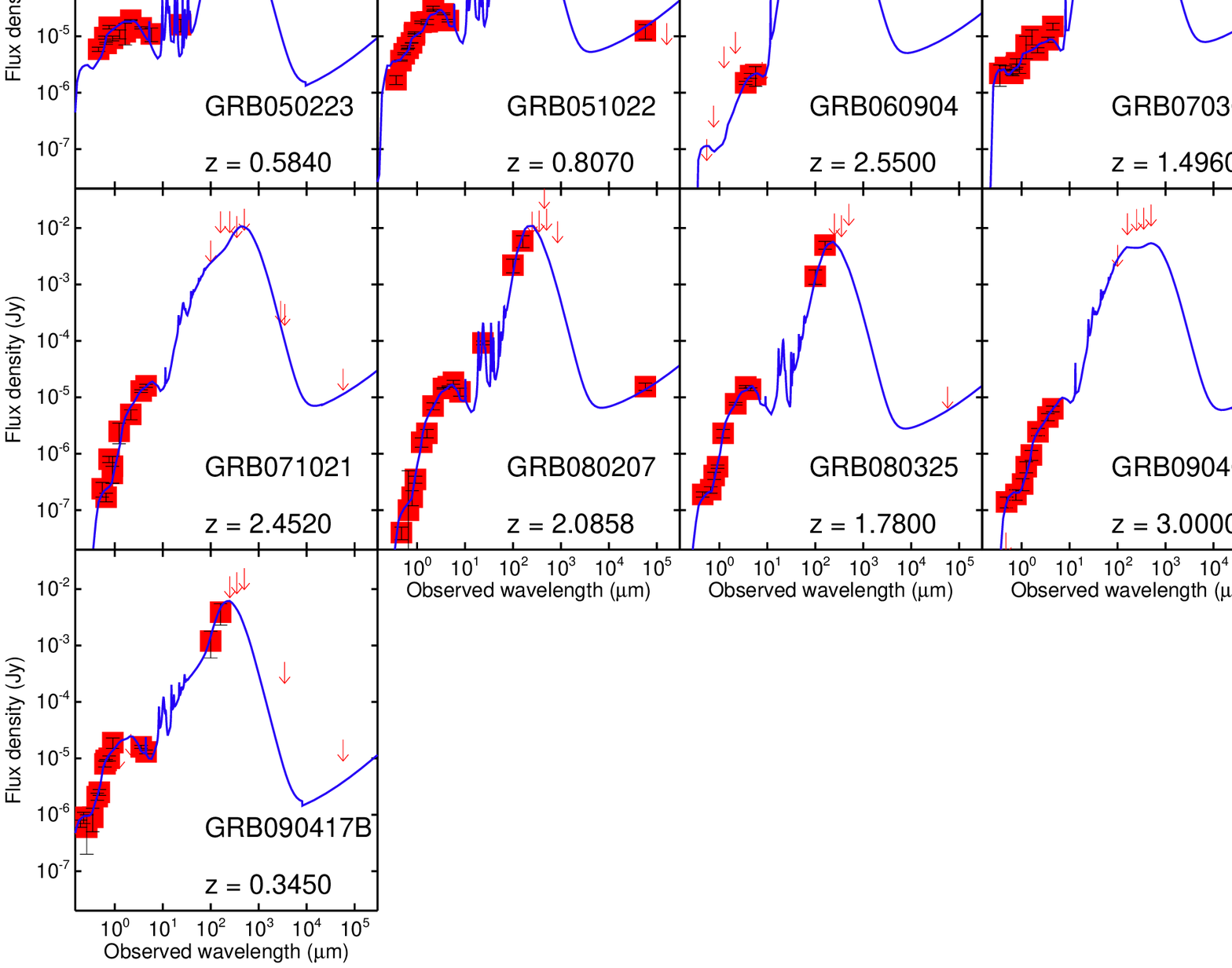}
}
\medskip
\caption{Best-fitting \grasil\ models in Jy plotted against wavelength,
superimposed on the multi-wavelength data for the GRBHs. 
Upper limits are shown as downward arrows where the head
of the arrow indicates the 3$\sigma$ upper limit.
}
\label{fig:grasil}
\end{figure*}

\subsection{{\it Herschel} photometry }
\label{sec:phot}

We checked the astrometry of the \hers\ images using astrometry of USNO
stars in the field, and translated the \hers\ images
when necessary to be consistent with the \spit/IRAC astrometry.
In each PACS and SPIRE image, an estimate for the background was obtained by
averaging the flux measured within a set of empty sky apertures close to the galaxies.
After background subtraction, the flux densities of the entire host galaxy at
all \hers\ wavelengths were obtained in apertures of 
radius 6\arcsec, except for the host of GRB\,020819B for which we
used an aperture of radius 13\arcsec. 
This larger aperture was used because of the large angular size
of the GRB\,020819B host \citep[][]{levesque10b}.
Following the PACS calibration guidelines\footnote{These are found
at the URL: http://herschel.esac.esa.int/twiki/
bin/view/Public/PacsCalibrationWeb\#PACS\_calibration\_and\_performance.},
we adjusted the photometry for the appropriate aperture
correction, and corrected the uncertainty estimates for correlated noise. 
Color corrections are 
around unity within the uncertainties, so we
neglected them.
The uncertainties associated with the measured photometry
were computed as a combination in quadrature of the
calibration uncertainty, 7\% for SPIRE data (according to Version 2.4, 2011
June 7, of the SPIRE Observer's Manual) and 5\% for PACS data (according to
Version 2.3, 2011 June 8 of the PACS Observer's Manual), and the sky
uncertainty derived by considering the number of pixels of the given aperture
and the standard deviation of the average value in the
individual sky apertures. Before addition in quadrature,
the sky uncertainties were corrected for the correlated noise, since the pixels
are not independent.  
The \hers\ photometry and its uncertainties are given in
Appendix \ref{app:tables}, Table~\ref{tab:photometry}.

We detect 7 GRBHs with \hers\ at redshifts ranging from $z=0.35-2.1$; all of these detections
are PACS, but GRBH\,020819B ($z=0.4$) also has SPIRE detections.
Figure \ref{fig:herschel} shows the PACS\,100\,\micron\ images for the 7
detections overlaid on IRAC images (3.6\,\micron\ except for the host of
GRB\,980613 which is 4.5\,\micron).
The (0,0) position corresponds to the host coordinates reported in Table \ref{tab:sample},
and is marked by a $+$. 
The host of GRB\,080325 has only a marginal detection, significant at $\sim 2.5\sigma$.

\subsection{Other multi-wavelength data}
\label{sec:otherdata}

To complete the SEDs, we gathered 
broadband optical, near-infrared, millimeter and centimeter portions of the SED 
from our own datasets or from the literature. 
In particular,
we obtained new optical/near-infrared data for the hosts of GRBs\,050219A and 050223
with the Gamma-Ray burst Optical and Near-infrared Detector (GROND)
\citep{greiner08}.
These data were reduced in a standard manner, using mainly
the GROND pipeline \citep{yoldas08,kruhler08}.
Aperture photometry was performed by using an aperture twice the diameter
of the full-width half-maximum of the stellar point-spread function (PSF).

As described in Sect. \ref{sec:sample}, 
we retrieved the infrared data for all the hosts in our sample from the \spit\ archive,
and reduced the available IRAC and MIPS data with MOPEX \citep{mopex}.
We then performed photometry on these images 
using a small circular aperture with a radius of 4\,\arcsec\ 
($\sim$5-7 times the IRAC PSF), 
with background subtraction determined from empty sky regions
around the source.
Aperture corrections have been applied only in a few cases
(the hosts of GRBs\,020127, 020819B, 050219A, and 080207), in order to 
take into account for their extension relative to the small apertures.
For the IRAC images of GRBH\,090404, because of the possible contamination
from the nearby galaxies \citep[see][]{perley13}, we applied the procedure
described by \citet{molinari11} to extract the host fluxes.

Radio and submm data were taken from 
\citet{berger01,berger03,tanvir04,stanway10,deugarte12,hatsukade12,svensson12,perleyperley13}.
The photometry used for the SED fitting is reported in Appendix \ref{app:tables}
(Table \ref{tab:photometry}), together with additional references. 

\section{Fitting the spectral energy distributions}
\label{sec:grasil}

We have used the multi-wavelength dataset described in the previous section
to estimate the dust masses, stellar masses, and SFRs in the sample GRBHs.
For the first time, we are able to 
place constraints on the dust emission in a significant
sample of GRBHs.
Before fitting the SED, photometry was corrected for Galactic extinction
assuming the values of \av\  taken from the NASA Extragalactic Database,
(NED, {\tt http://ned.ipac.caltech.edu}), 
and using the interstellar extinction curve by \citet{cardelli89}.

We applied the SED fitting method introduced by \citet[][]{michalowski09,michalowski10}
based on 35\,000 templates in the library of \citet{iglesias07}, plus
additional templates of \citet{silva98} and \citet{michalowski08}, all
developed in \grasil\footnote{\url{www.adlibitum.oat.ts.astro.it/silva}}
\citep{silva98}. They are based on numerical calculations of radiative transfer
within a galaxy, which is assumed to be a triaxial system with diffuse dust and
dense molecular clouds that are the sites of star formation.
A discussion of the derivation of galaxy properties and typical
uncertainties is given by \citet{michalowski09}.

\begin{center}
\begin{table*}[!ht]
      \caption[]{\grasil\ fitting results}
\label{tab:grasil}
\resizebox{\linewidth}{!}{
{\small
\begin{tabular}{lrrlrcrrrrc}
\hline
\multicolumn{1}{c}{GRB Host}  & 
\multicolumn{1}{c}{SFR(UV)$^\mathrm{a}$} & 
\multicolumn{1}{c}{SFR(IR)$^\mathrm{a}$} & 
\multicolumn{1}{c}{\av} & 
\multicolumn{1}{c}{Log} & 
\multicolumn{1}{c}{\hers} & 
\multicolumn{1}{c}{Log} &
\multicolumn{1}{c}{Log} &
\multicolumn{1}{c}{Log} &
\multicolumn{1}{c}{T$_{\rm dust}$} & 
\multicolumn{1}{c}{Modified blackbody fits} \\ 
\multicolumn{1}{c}{galaxy}  & 
\multicolumn{1}{c}{(\sfr)}  & 
\multicolumn{1}{c}{(\sfr)}  & 
\multicolumn{1}{c}{(mag)}  & 
\multicolumn{1}{c}{\lir}  & 
\multicolumn{1}{c}{Detected?}  & 
\multicolumn{1}{c}{M$_{\rm stars}^\mathrm{a}$}  & 
\multicolumn{1}{c}{M$_{\rm burst}^\mathrm{a}$}  & 
\multicolumn{1}{c}{M$_{\rm dust}$}  & 
\multicolumn{1}{c}{(\grasil)}  & 
\multicolumn{1}{c}{Log(M$_{\rm dust}$)} \\
&&&&\multicolumn{1}{c}{(\lsun)}  & &
\multicolumn{1}{c}{(\msun)}  & 
\multicolumn{1}{c}{(\msun)}  & 
\multicolumn{1}{c}{(\msun)}  & 
\multicolumn{1}{c}{(K)}  & 
\multicolumn{1}{c}{(\msun)}  \\ 
\multicolumn{1}{c}{(1)}  & 
\multicolumn{1}{c}{(2)}  & 
\multicolumn{1}{c}{(3)}  & 
\multicolumn{1}{c}{(4)}  & 
\multicolumn{1}{c}{(5)}  & 
\multicolumn{1}{c}{(6)}  & 
\multicolumn{1}{c}{(7)}  & 
\multicolumn{1}{c}{(8)}  & 
\multicolumn{1}{c}{(9)}  & 
\multicolumn{1}{c}{(10)}  & 
\multicolumn{1}{c}{(11)}  \\ 
\hline
\\
970828  &    0.3 &   14.4 &   2.38 &  11.18 &  No &   9.80 &   8.96 & $-$ &   38.7 & $<$8.21   \\
980613  &    1.5 &   39.5 &   1.94 &  11.62 & Yes &   9.31 &   9.28 &   7.53 &   71.5 & $-$ \\
980703  &    2.7 &   86.4 &   1.94 &  11.96 &  No &   9.98 &   9.60 & $-$ &   38.7 & $<$8.65   \\
990705  &    2.3 &    6.6 &   0.16 &  10.84 &  No &  10.50 & $-$    & $-$ &   28.5 & $<$8.18   \\
020127  &    2.7 &   56.3 &   0.39 &  11.77 &  No &  11.51 &   9.54 & $-$ &   61.3 & $<$9.01   \\
020819B &    3.5 &   13.0 &   0.54 &  11.13 & Yes &  10.52 & $-$    &   8.99 &   24.4 & $-$ \\
030115  &    1.6 &  159.1 &   1.95 &  12.22 &  No &  10.87 &   9.85 & $-$ &   33.2 & $<$9.53   \\
050219A & $<$0.1 & $<$0.1 &   0.00 &   8.29 &  No &   9.91 & $-$    & $-$ &   47.2 & $<$6.37   \\
050223  &    1.9 &    1.7 &   0.21 &  10.25 &  No &  10.02 & $-$    & $-$ &   21.0 & $<$7.87   \\
051022  &    4.8 &   17.9 &   0.56 &  11.27 & Yes &  10.29 &   9.28 &   7.27 &   52.6 & $-$ \\
060904A  &    1.1 &  303.9 &   4.17 &  12.50 &  No &  10.18 &  10.17 & $<$8.41   &  132.0 & $-$ \\
070306  &    9.6 &  144.1 &   1.65 &  12.18 & Yes &  10.05 &   9.97 &   8.29 &   52.6 & $-$ \\
071021  &    2.6 &  288.2 &   1.72 &  12.48 &  No &  11.40 &  10.15 & $-$    &   33.2 & $<$9.50   \\
080207  &    1.0 &  170.1 &   1.69 &  12.25 & Yes &  11.17 &  10.16 &   8.15 &   61.3 & $-$ \\
080325  &    1.2 &   66.5 &   1.03 &  11.84 & Yes &  11.09 &   9.62 &   8.06 &   52.6 & $-$ \\
090404  &    3.1 &  380.8 &   2.28 &  12.60 &  No &  11.10 &  10.30 & $-$    &   52.6 & $<$9.93   \\
090417B &    0.2 &    1.7 &   0.54 &  10.25 & Yes &   9.73 & $-$    &   8.18 &   21.0 & $-$ \\
\\
\hline
\end{tabular}
}
}
\vspace{0.5\baselineskip}
\begin{description}
\item
[$^{\mathrm{a}}$] 
The SFRs and stellar masses have been converted to a \citet{chabrier03} IMF
by dividing by a factor of 1.8.
\item
[Col. (2)]=\,SFR inferred from SED fitting and rest-frame UV,
uncorrected for extinction;
\item
[Col. (3)]=\,SFR inferred from the IR emission
from radiation reprocessed by dust;
\item
[Col. (4)]=\,total \av\ obtained by considering the total
attenuation in both the diffuse ISM and the dense molecular cloud
components as described in the text;
\item
[Col. (5)]=\,total IR luminosity from dust emission;
\item
[Col. (7)]=\,total stellar mass of the galaxy;
\item
[Col. (8)]=\,stellar mass of the burst episode assumed to
have onset 50\,Myr ago;
\item
[Col. (9)]=\,total dust mass of the galaxy;
\item
[Col. (10)]=\,luminosity-weighted mean dust temperature as given by \grasil;
\item
[Col. (11)]=\,the fixed-temperature ($T$\,=\,35$\pm$1\,K)
fixed-emissivity ($\beta$\,=\,2.0)
MBB fits are given when the \grasil\ fit
does not give a realistic upper limit.
\end{description}
\end{table*}
\end{center}

The templates cover a broad range of galaxy properties from quiescent to
starburst. 
Their star formation histories (SFHs) are assumed to be a smooth 
Kennicutt-Schmidt (KS)-type law
\citep[SFR proportional to the gas mass to some power, see][for
details]{silva98} with a starburst (if any) on top of that starting $50$ Myr
before the current epoch at that redshift. There are seven free parameters
in the library of \citet{iglesias07}: the normalization of the KS
law, the timescale of the mass infall, the intensity of the starburst, the
timescale of the molecular cloud destruction, the optical depth of molecular
clouds, the age of the galaxy and the inclination of the disk with respect to the
observer.
These templates are based on a Salpeter \citep{salpeter55}
Initial Mass Function (IMF), but in the analysis
the stellar masses and SFRs have been converted to a \citet{chabrier03} IMF 
by dividing by a factor of 1.8. 

\subsection{Uncertainties on the fitted parameters}
\label{sec:uncertainties}

SED fitting procedures are notoriously degenerate \citep[e.g.,][]{leborgne02},
especially when estimating photometric redshifts (as in
GRBH\,030115, see Table \ref{tab:sample}).
We have therefore run a series of tests to establish the realistic
uncertainties on the various parameters determined by \grasil.
These tests have shown that the stellar masses are formally good to within 40\%,
but comparison with other templates (see Sect. \ref{sec:comparison} below) suggests
that a factor of 2 may be more realistic when comparing with other samples.
The stellar portion of the SED for our hosts is usually well constrained, but
stellar masses tend to differ according to the adopted SFHs \citep[see][]{michalowski12a}.
SFR values (both UV, IR)
are accurate to roughly 30\%, and the burst masses to within a factor of 2.
The dust masses have an uncertainty of $\sim$0.3\,dex (a factor of 2), and the
dust temperature 10$-$20\,K, 
or even more because of the sparse sampling of the SED around its peak.

\subsection{Modified blackbody fits}
\label{sec:mbb}

In order to obtain realistic upper limits to the dust mass, 
in the cases where \grasil\ models were unable to do so because of non-detections,
we also fit single-temperature modified blackbodies
(MBBs) to the IR SED using PACS upper limits.
Such fits assume that the dust is optically thin, with an emissivity that varies
as a power law, with index $\beta$.
These are very simplistic fits, in which we attempted only to derive the maximum dust
mass that would pass through the data points.
The MBB emissivity index $\beta$ was fixed to 2 \citep[e.g.,][]{bianchi13},
and the temperature was fixed to $T\,=\,35\pm1$\,K \citep[e.g.,][]{michalowski08,michalowski10}.
The dust mass was calculated using the Milky Way dust opacities given by
\citet{draine07}.
The upper limit to the dust mass was taken to be the inferred total dust mass
calculated at 160\,\micron.
In only one case, the host of GRB\,060904A, is the \grasil\ approximation of the non-detections
more realistic.
We repeated the procedure also for the detected GRBHs, and find agreement 
with the \grasil\ values to within a factor of 2.
Hence, the MBB upper limits are expected to be compatible with \grasil, and can
be analyzed in a coherent way.

\section{Results}
\label{sec:results}

Figure \ref{fig:grasil} shows the best-fit \grasil\ models of the 17 GRBH
multiwavelength SEDs.
Table \ref{tab:grasil} reports the best-fit parameters given by \grasil.
In the cases where there is no \hers\ detection 
(see Col. 6 of Table \ref{tab:grasil}), the dust masses
are intended to be upper limits (as noted by $<$).
Col. 11 of Table \ref{tab:grasil} gives the resulting \mdust\ for the MBB
fitting as described in Sect. \ref{sec:mbb}.
These are intended only as very approximate upper limits to the dust mass,
given the flux limits implied by our \hers\ observations.
The values of \av\ given by the \grasil\ fits are an average over the entire galaxy,
and take into account the total attenuation given by the extinction in the diffuse
interstellar medium (ISM) combined with that in the dense molecular clouds
\citep[see][]{silva98}.

Unlike GRB samples that are selected in order to be as 
statistically unbiased as possible
\citep[][]{fynbo09,greiner11,hjorth12,salvaterra12},
our sample of GRBHs is not a statistically complete sample in any sense.
Nevertheless, because our sample is dominated by hosts of dark GRBs
(14/17), it is unique and,
in combination with other samples from the
literature \citep[e.g.,][]{savaglio09,perley13}\footnote{In the \citet{savaglio09} sample,
as noted before, we have eliminated from further consideration the hosts of 6 short GRBs
studied therein.},
can give a new perspective on the characterization of the 
galaxy masses, dust content, and SFR of the long-GRBH population.
After comparing our results with previous work, in the
following sections we compare the stellar masses, dust masses,
SFRs, and host extinctions of the combined sample with other galaxy
populations covering a similar redshift interval.
We postpone the discussion of trends of stellar mass and
SFR with redshift to Sect. \ref{sec:context}.

\begin{figure}[!h]
\centerline{
\includegraphics[angle=0,height=0.85\linewidth,bb=18 160 592 650]{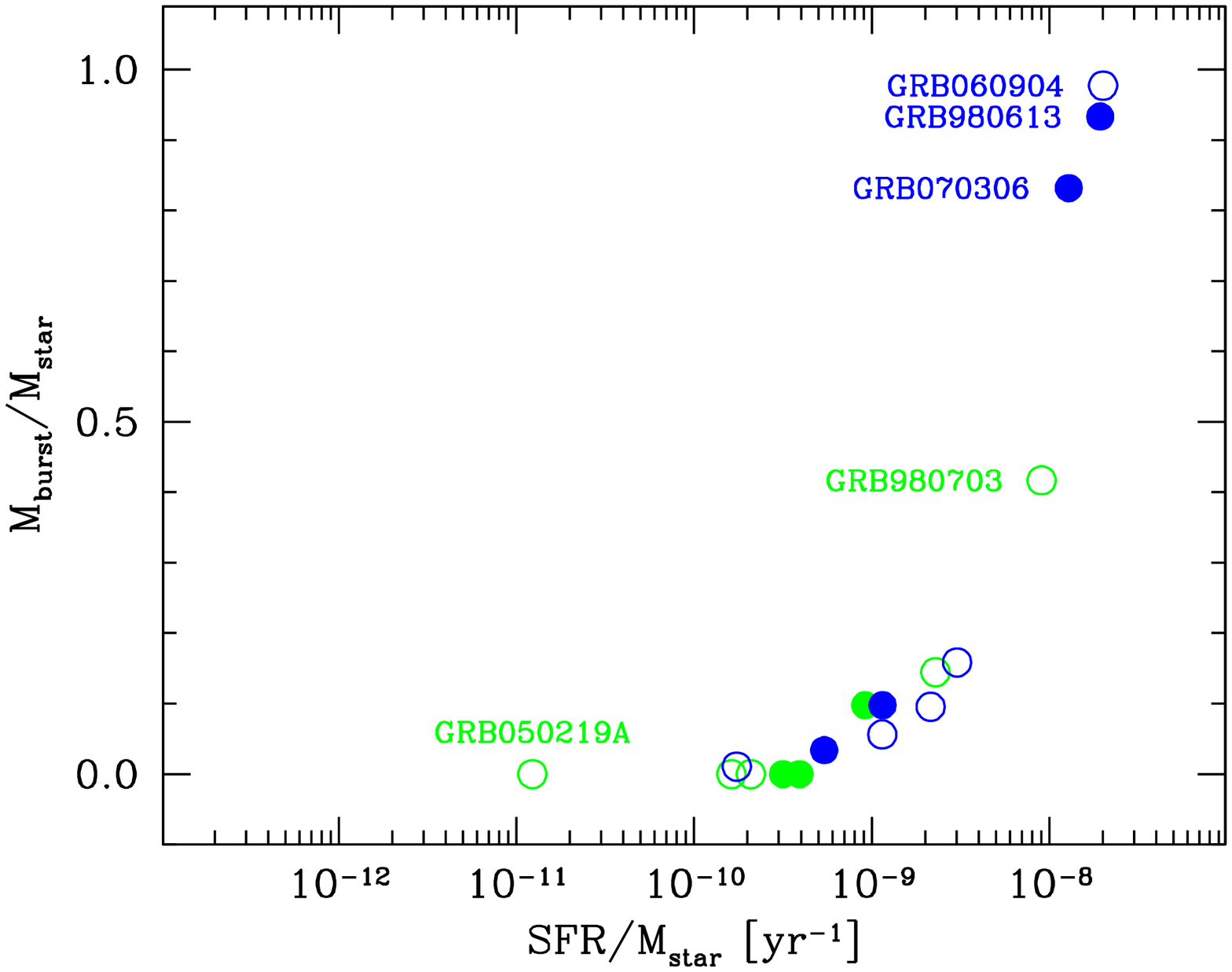}
}
\caption{Ratio of stellar mass in burst M$_{\rm burst}$ to total stellar mass M$_{\rm star}$
vs. specific SFR.
\hers\ detections are shown with filled symbols, and non-detections with open ones;
blue symbols correspond to GRBHs with $z>$1.1 (\zmed)
and green ones to GRBHs with $z\leq$1.1.
The most extreme hosts are labeled by the GRB name.
}
\label{fig:burst}
\end{figure}

\subsection{Comparison with previous work}
\label{sec:comparison}

Because of the lack of detections at IR or sub-mm wavelengths
\citep[e.g.,][]{berger03,tanvir04,priddey06},
dust masses have been determined for very few GRB hosts before now \citep[e.g.,][]{michalowski08}.
The dust mass for one of our \hers\ targets, GRBH\,980703, was also 
calculated by \citet{michalowski08}, and our fit here is very similar
to that found with earlier data (no IR detections then or now), 
giving virtually identical limits to
the dust mass.

Stellar masses can be compared more readily.
Six galaxies in our sample have been studied by \citet{savaglio09}
or \citet{castro10}, and 12 are in common with \citet{perley13}.
To estimate stellar masses,
\citet{castro10} uses rest-frame $K$-band luminosities with an approximate
mass-to-light ratio; correcting both mass estimates to a \citet{chabrier03} IMF
gives a mean ratio of {dex($-0.21$)}\,\msun, 
with our masses being $\sim$1.6 times smaller than theirs.
\citet{savaglio09} use SED fitting to the rest-frame optical and UV photometry,
and consider two star-formation episodes; their stellar masses are, in the mean,
2.5 times smaller than ours, perhaps because of the lack of \spit/IRAC
data which could be important for our dark-GRB dominated sample.
\citet{perley13} use SED fitting with a single episode of star formation,
but include \spit/IRAC data where possible.
Our estimates of stellar masses are, on average, 
1.7 times larger than theirs, with a 
large standard deviation.
One possible cause of this difference is that \grasil\ uses
a two-episode SFH which tends to give larger stellar masses
than estimates based on one star-formation episode \citep{michalowski12a}.
The biggest discrepancy 
is for the host of GRB\,980613, for which the estimate by \citet{castro10}
is roughly 40 times larger than ours, and $\sim$200 times larger than the
one given by \citet{perley13}. 
This could be due to the complex configuration of this host,
which consists of at least five galaxies, or galaxy fragments \citep{djor03}. 
In general, the scatter among all the comparisons is 
$\sim$dex(0.4$-$0.6), or roughly a factor of 3.
Given the different approaches and the different sets of photometry,
this could be considered reasonable agreement,
and probably a realistic estimate of the uncertainty. 

The comparison of the stellar masses of the 20 galaxies (not in our sample) common to 
\citet{savaglio09} and \citet{perley13} show very good agreement between these two
studies, with
a mean difference of dex(0.03) and a standard deviation between the two sets
of measurements of dex(0.4), or a factor of $\sim$2.5.
We consider this to be good agreement, and in what follows will include
both samples in our analysis when possible.

We have compared our SFRs 
to those found by \citet{savaglio09}, \citet{castro10} and \citet{perley13}, and 
find similar agreement.
Again, correcting all SFR estimates to the \citet{chabrier03} IMF,
we find that our estimates estimated from the UV, SFR(UV), are, in the mean, 
within a few percent of those by \citet{castro10} (with a spread of roughly
a factor of 2).
Because neither the \grasil\ SFR(UV)s 
nor the \citet{castro10} estimates have been corrected for extinction,
this good agreement is probably an indication that 
extinction corrections are the biggest stumbling block for consistent
UV-derived SFRs.

The comparison with \grasil\ SFRs estimated from the IR, SFR(IR), is less straightforward.
It is well known that IR-inferred SFRs can be much larger than those
derived from fitting the UV continuum \citep[e.g.,][]{berger03,buat10}.
For our sample, the mean ratio of SFR(IR)/SFR(UV)
given by the \grasil\ fits is $\sim$16.
For the 12 galaxies in common with \citet{perley13},
their SFRs are on average $\sim$8 times larger than 
the \grasil\ UV-based SFRs (uncorrected for extinction)
and roughly half the IR SFRs given by our \grasil\ fits. 
For the six galaxies in common with \citet{savaglio09}, the \grasil\ SFR(UVs)
are $\sim$3 times smaller than theirs and the SFR(IR) $\sim$3 times larger.
Given the different methods used to derive SFRs (emission lines, SED fitting), 
such discrepancies are not unexpected.
The SFRs given by \citet{savaglio09} and \citet{perley13} lie between the SFR(IR) and 
SFR(UV) extremes, although their greater similarity with SFR(IR) from \grasil\ implies that 
careful SED multi-wavelength fitting with reasonable assumptions for the SFH
can partially compensate the lack of IR or sub-mm data. 
We conclude that the SFRs we obtain here
are reasonable estimates, and in what follows we will adopt the 
IR-derived SFRs enabled by our \hers\ observations.

\begin{figure*}[!ht]
\centerline{
\includegraphics[angle=0,height=0.48\linewidth,bb=18 160 592 650]{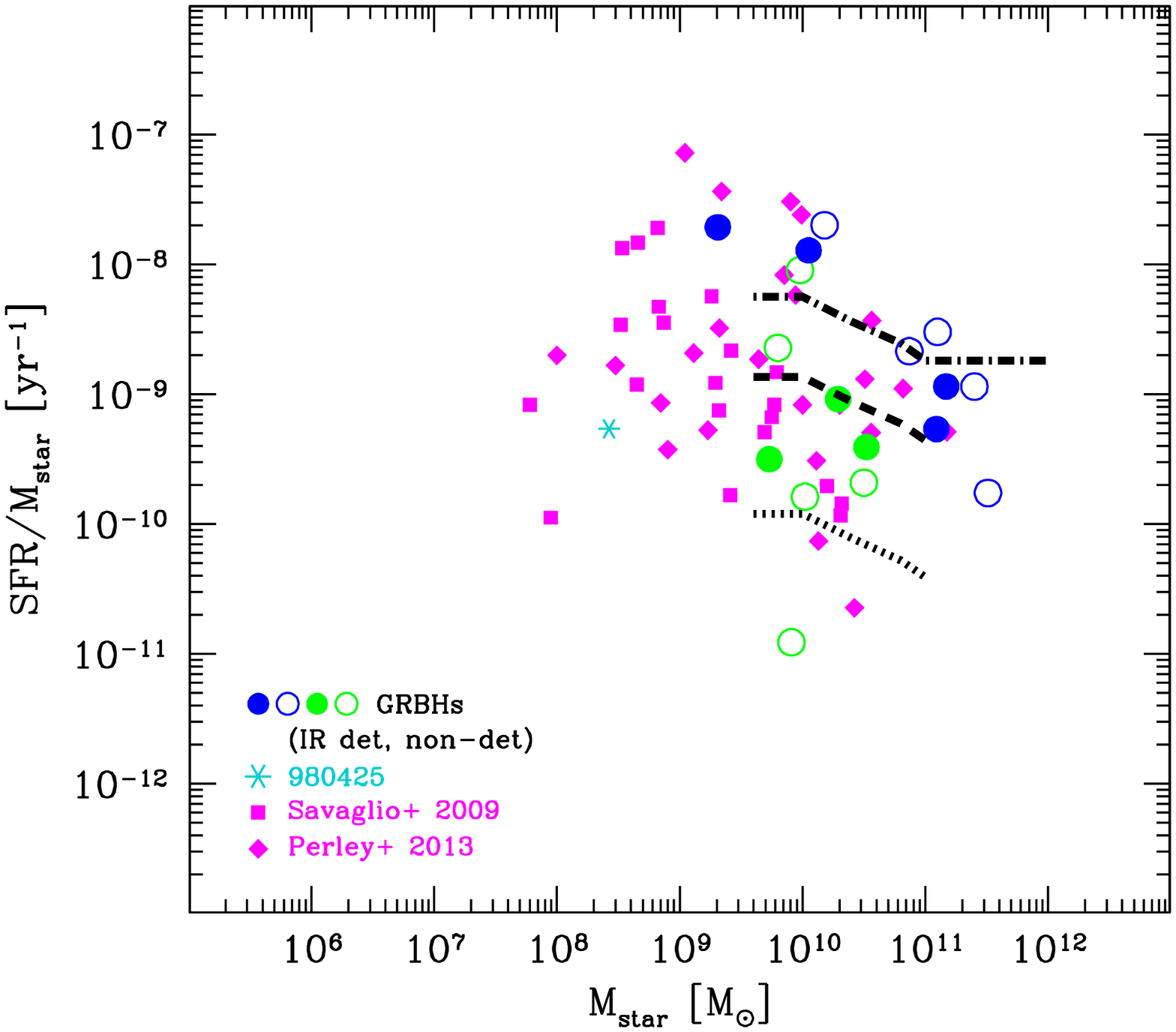}
\hspace{-2.0cm}
\includegraphics[angle=0,height=0.48\linewidth,bb=18 160 592 650]{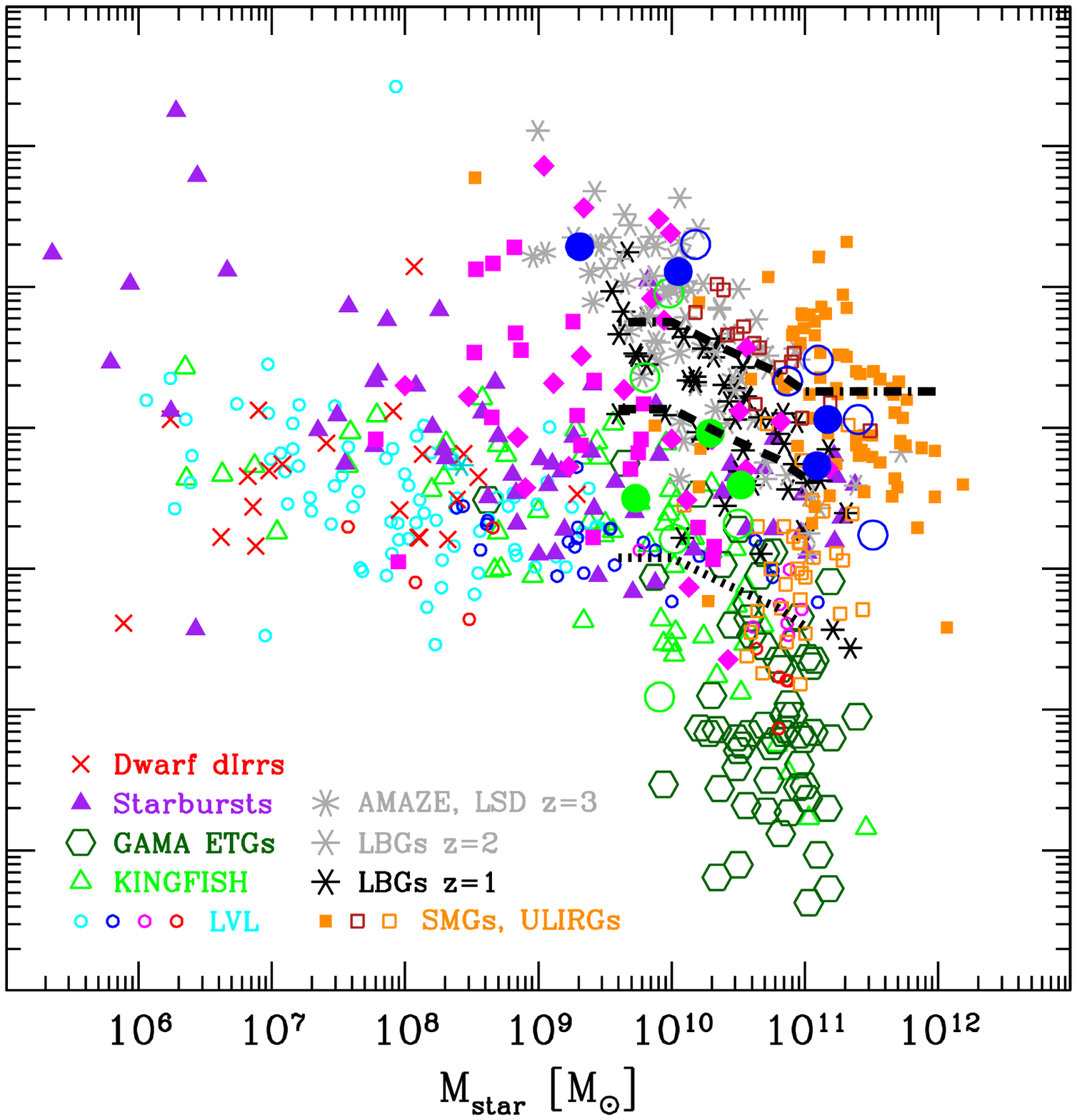}
}
\captionof{figure}{Specific SFR plotted against stellar mass:
the left panel shows GRBHs only 
and the right panel includes other galaxy populations.
The three curves are the fits to the star-formation MS at $z=0.2-0.4$ (dotted lower),
$z=1.0-1.2$ (dashed middle), and $z=2.0-2.5$ (dot-dashed upper) by \citet{karim11}.
The GRBHs observed by \hers\ are indicated by 
filled (for IR detections) and open (IR non-detections) circles;
blue symbols correspond to GRBHs with $z>$1.1 (median $z$ of the sample),
and green ones to GRBHs with $z\leq$1.1.
GRBHs from \citet[][excluding duplicates with our sample]{perley13,savaglio09} are shown
as filled (magenta) diamonds and as filled (magenta) squares, respectively.
The host of GRB\,980425 is shown with (cyan) 6-pronged asterisks \citep{michalowski13}.
Star-forming galaxies at $z\sim0$ from the KINGFISH sample are shown as open (light green) triangles
\citep{kennicutt11} and GAMA early-type galaxies (ETGs) from \citet{rowlands12} by
open (dark green) hexagons.
The parameters for the other $z\sim0$ galaxy populations have been taken from \citet{hunt12}
with galaxies in the Local Volume Legacy (LVL) survey shown as open circles,
color-coded for Hubble type (red are early types, cyan are late types);
dwarf irregular galaxies at $z\sim0$ are shown as (red) $\times$; 
starbursts at $z\sim0$ as (purple) filled triangles. 
$z\sim2$ SMGs and $z\sim1-2$ ULIRGs from \citet{michalowski10} and \citet{lofaro13}
are given by (orange) filled and open squares, respectively;
(firebrick) open squares correspond to $z\sim0$ ULIRGs 
from \citet{dacunha10a}.
Lyman-break galaxies (LBGs) at $z\sim1$ and $z\sim2$ are indicated by 6-pronged asterisks 
\citep{shapley04,shapley05,oteo13a,oteo13b};
LBGs at $z\sim3$ by 8-pronged asterisks \citep{maiolino08,mannucci09}.
}
\label{fig:ms}
\end{figure*}

\subsection{Ages and star-formation histories of GRBHs}
\label{sec:age}

The \grasil\ fits give us some idea of what the SFH of the hosts has been.
In particular, we can understand if a recent burst of star formation (over
the last $\sim$50\,Myr, see Sect. \ref{sec:grasil})
has contributed significantly to the stellar mass budget, or whether
most of the stars have been created over longer timescales.
In Fig. \ref{fig:burst}, we plot against specific SFR (sSFR, SFR/\mstar)
the ratio of mass produced by the recent starburst episode \mburst\
and the total stellar mass in the galaxy, \mstar.
The SFR considered in the sSFR is that inferred by \grasil\ from the IR luminosity and corresponds
to the star formation over the last several hundreds of Myr, rather than the most
recent burst.
Although the quantities in the two axes are highly correlated,
the comparison shows that the SFH calculated by \grasil\ is realistic,
or at least self-consistent with the SFR and \mstar.

Fig. \ref{fig:burst} illustrates the expected result that the hosts with the
highest sSFR also have a significant fraction of their stellar mass produced
in a recent burst; half the galaxies in the observed sample have $\ga$10\%
of their stellar mass from such an episode.
There is clearly a trend of \mburst/\mstar\ with  sSFR; the
four galaxies with the highest ratio are also those with
sSFR$\ga$10$^{-8}$\,yr$^{-1}$, and would be considered starbursts
at any redshift \citep[at least $z\la$\,3, e.g.,][]{elbaz11}.
These are also among the galaxies with the least pronounced rest-frame 1.6\,\micron\
photometric bump (see Fig. \ref{fig:grasil})
generally associated with evolved stellar populations.
Because of increased stellar opacities,
the strength of this bump increases with
age and metallicities \citep{simpson99,sawicki02}, so a weak 
1.6\,\micron\ feature signifies young age or sub-solar metallicity (or both).
Because of the correlation we find between \mburst/\mstar\ and sSFR,
we attribute the weakness of this bump in the hosts of GRBs\,980613,
980703, 060904A, and 070306 to young ages, although these galaxies may 
be relatively metal poor as well.

Interestingly, the \hers\ detections are not uniquely associated with high sSFRs
($\ga10^{-8}$\,yr$^{-1}$).
Of the four shown in Fig. \ref{fig:burst} 
only GRBHs\,980613 and 070306 are detected with PACS. 

The GRBH with the lowest sSFR ($\sim10^{-11}$\,yr$^{-1}$)
is the host of GRB\,050219A (see also Fig. \ref{fig:burst}), 
and it is also the
galaxy with the smallest amount of dust and the lowest SFR (see Table \ref{tab:grasil}).
This galaxy is peculiar for a host of a long GRB, and will be discussed in a future
paper \citep{rossi14}.


\begin{figure*}[!ht]
\centerline{
\includegraphics[angle=0,height=0.48\linewidth,bb=18 160 592 650]{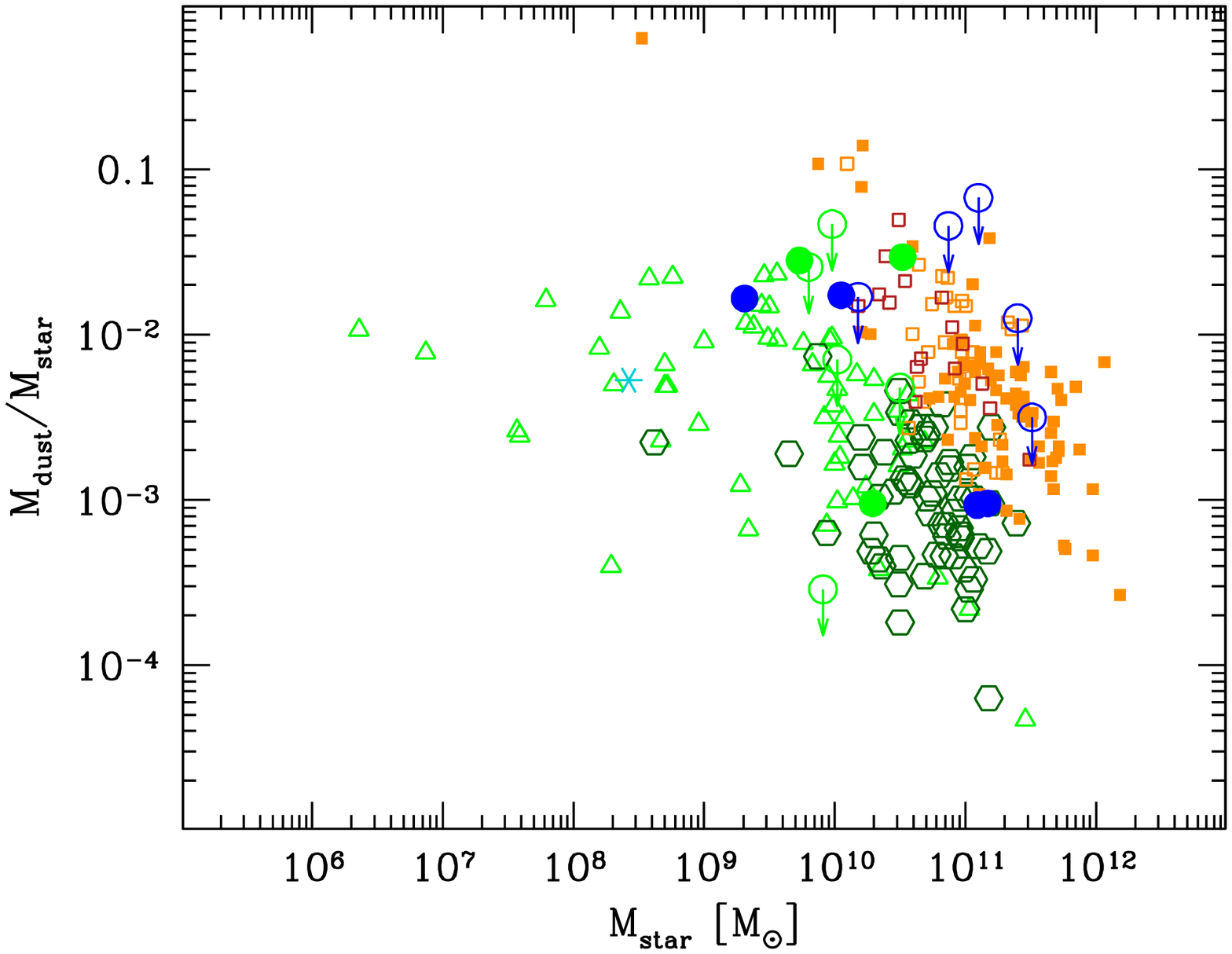}
\hspace{-1.9cm}
\includegraphics[angle=0,height=0.48\linewidth,bb=18 160 592 650]{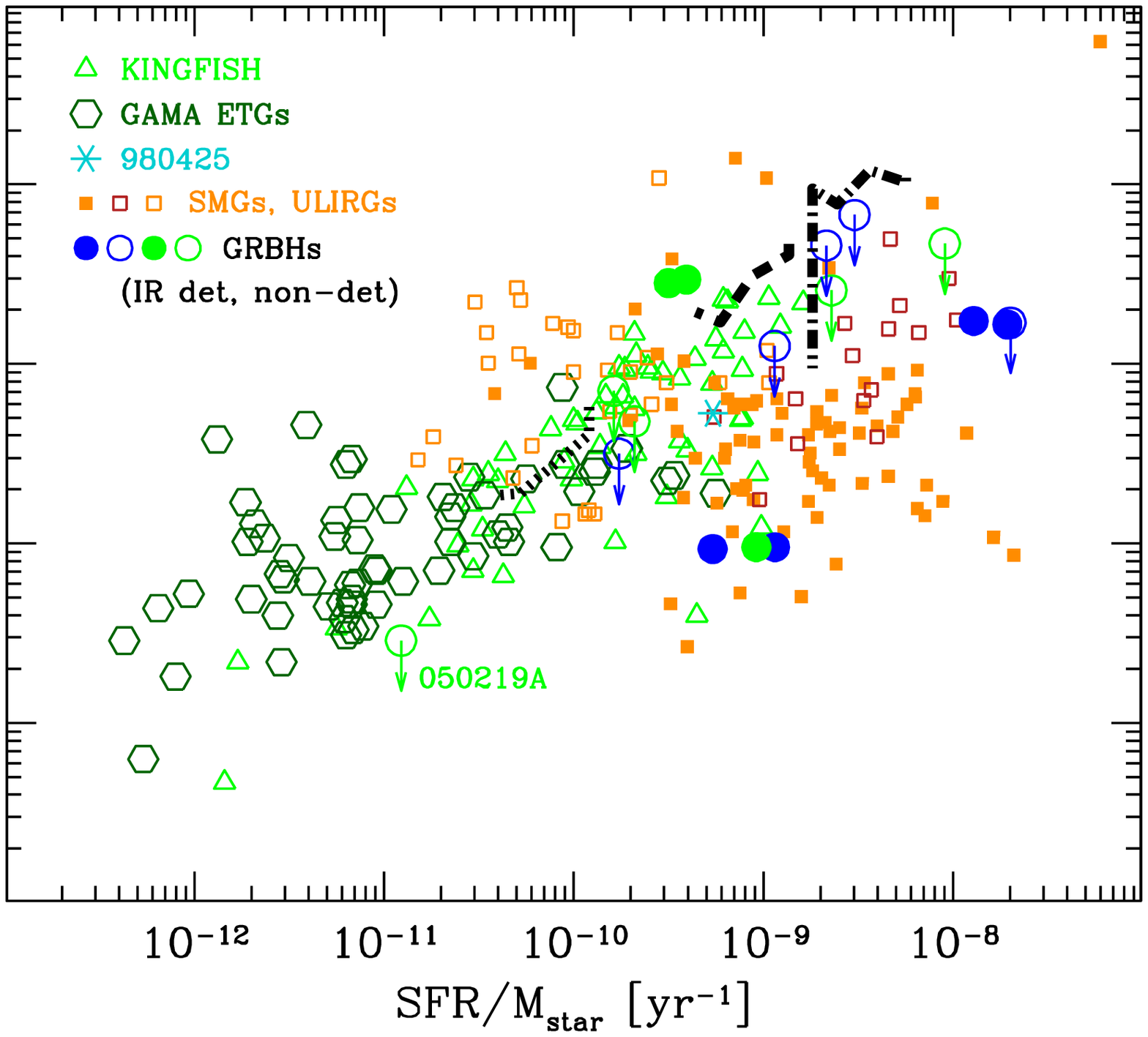}
}
\caption{Ratio of \mdust/\mstar\ plotted against stellar mass \mstar\ (left panel)
and against SFR/\mstar\ sSFR (right).
As in Fig. \ref{fig:ms},
the GRBHs observed by \hers\ are indicated by 
filled (for IR detections) and open (IR non-detections) circles,
and the other galaxy samples shown are also coded as in Fig. \ref{fig:ms}.
The three curves in the right panel are the fits of sSFR vs. \mstar\
at $z=0.2-0.4$ (dotted lower), $z=1.0-1.2$ (dashed middle), 
and $z=2.0-2.5$ (dot-dashed upper) by \citet{karim11},
as shown in Fig. \ref{fig:ms}, but here incorporating the dependence of \mdust\ on SFR
as found at $z\approx0$ by \citet{dacunha10b}.
The vertical trend of the $z\sim2$ curve is related to the fall-off of sSFR
at high stellar masses (see Fig. \ref{fig:ms}).
}
\label{fig:duststars}
\end{figure*}

\subsection{Stellar masses and SFR}
\label{sec:mainsequence}

We explore in Fig. \ref{fig:ms} sSFR vs. \mstar, or the ``main sequence'' (MS) of
star formation \citep[e.g.,][]{salim07,noeske07}. 
To augment the GRBH statistics, in this and subsequent figures when possible,
we have included the 23 hosts identified with {\it Swift} GRBs and
the 31 pre-{\it Swift} ones studied by \citet{perley13}, together with
the hosts of 40 {\it long} mostly optically-bright GRBs studied by \citet{savaglio09}.
These numbers do not exclude the galaxies in common with our study (see Sect. \ref{sec:context}
for a fuller discussion).
In the rest of the paper, where there are duplicates, we will prefer
our values of \mstar\ and SFR, then those of \citet{perley13}, except where
the quality of the fit is poor, as given by reduced $\chi_\nu^2\leq4$; 
in this case we take the parameters from \citet{savaglio09}.
This order of preference is dictated by the inclusion of longer-wavelength data points in the 
SED fitting; even though the agreement between these two
data sets is quite good (see Sect. \ref{sec:comparison}), 
\citet{perley13} is preferred over the \citet{savaglio09} values 
because of the inclusion of \spit/IRAC data. 

The left panel of Fig. \ref{fig:ms} shows our sample as (green, blue for $z\leq$\zmed, $z>$\zmed,
respectively) circles, and galaxies 
from \citet{perley13} as filled (magenta) diamonds and from \citet{savaglio09} as filled (magenta) squares;
the right panel adds other galaxy populations from $z\sim0$ to $z\ga3$.
The host of GRB\,980425 \citep{michalowski13} is shown as a (cyan) asterisk;
its extremely low redshift ($z\,=\,0.0085$) makes it unique among GRBHs.
The curves in Fig. \ref{fig:ms} report the trends for the star-formation MS
fitted by \citet{karim11}
for $z\sim0$, $z\sim1$, and $z\sim2$ (from lower curve to the upper one).

It is clear from the figure that the GRBHs span a wide range in sSFRs,
and the most extreme objects have very high sSFRs, similar to or even more
extreme than the Lyman-break galaxies (LBGs) at $z\sim3$.
As expected, because higher-than-average sSFRs tend to be associated with 
lower-than-average stellar masses, such GRBHs are not the most massive.
The comparison with other galaxy populations and the trends of sSFR and \mstar\
expected with redshift illustrated by the curves in Figure \ref{fig:ms} suggest that the GRBHs
studied here are medium-size high-sSFR galaxies,
as might be expected for galaxies selected
on the basis of SFR because of the explosive GRB event.

\begin{figure}[!h]
\centerline{
\includegraphics[angle=0,height=0.9\linewidth,bb=18 144 592 650]{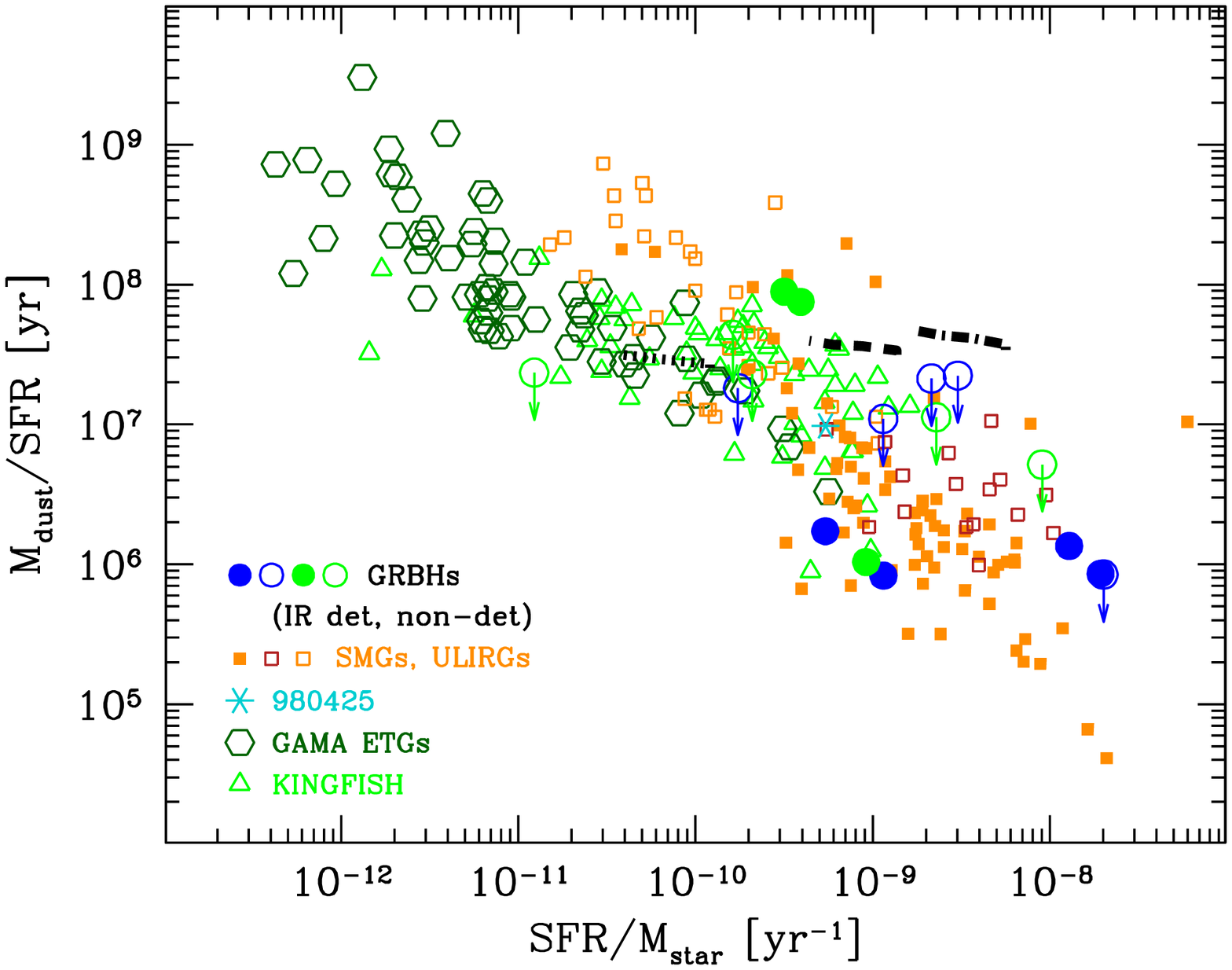}
}
\caption{Ratio of \mdust/SFR plotted against sSFR. 
As in Fig. \ref{fig:ms},
the GRBHs observed by \hers\ are indicated by  
filled (for IR detections) and open (IR non-detections) circles,
and the other galaxy samples shown are also coded as in Fig. \ref{fig:ms}.
The three curves 
are the fits of sSFR vs. \mstar\ at $z=0.2-0.4$ (dotted left), $z=1.0-1.2$ (dashed middle), 
and $z=2.0-2.5$ (dot-dashed right) by \citet{karim11}
as shown in Fig. \ref{fig:ms}, but here incorporating the dependence of \mdust\ on SFR
as found at $z\approx0$ by \citet{dacunha10b} (see also Fig. \ref{fig:duststars}).
}
\label{fig:dustsfr}
\end{figure}

\subsection{Dust, stars, and SFR}
\label{sec:dust}

Up to now, dust emission has been detected in very few GRBHs 
\citep{berger03,tanvir04,priddey06,wang12,lefloch12,michalowski13}. Pre-{\it Swift} studies showed 
a $\ga3\sigma$ detection rate at 850\,\micron\ of 2/26 hosts, $\sim$8\% \citep{tanvir04}.
Moreover, the blue colors of these brightest hosts at submm wavelengths 
\citep[e.g., GRB\,000418, GRB\,010222:][]{lefloch03,gorosabel03}
are not typical of galaxy populations selected in submm surveys. 
Of 15 GRBHs around $z\sim1$, only three were detected at 24\,\micron\ with \spit\
\citep[GRB\,970828, GRB\,980613, GRB\,990705:][]{lefloch06}.
These low detection rates of dust emission in GRBHs combined with measurements of stellar
mass and UV-inferred SFRs \citep[e.g.,][]{savaglio09}
led to the conclusion that the host population was biased against 
dusty, massive and strongly starbursting galaxies, at least for $z\la1.5$.

Over the last few years, however, evidence has emerged that GRBHs are a more diverse population
than previous work suggested.
Prompt follow-up observations have enabled the localization of afterglows in a significant
percentage of dark GRBs \citep[e.g.,][]{greiner11,melandri12,hjorth12}, 
thus opening the possibility of studying their hosts.
With our \hers\ observations, we have detected dust emission in 7 of 17 GRBHs, 6 of which
host dark GRBs. 
For dark GRBs alone, the detection rate is quite high, 6/14 (43\%).
The detection rate for hosts of optically-bright afterglows is lower (1/4, 25\%),
and since these are small numbers, the rate of optically-bright host detections
could be considered consistent with the previous, low, IR detection rates.
Although the statistics are sparse, the fractions suggest that
the hosts of dark GRBs are more
likely to be detected at IR/submm wavelengths than their optically-bright counterparts.

Recent work has shown that in star-forming galaxies
\mstar, \mdust, and SFR are mutually correlated \citep[e.g.,][]{dacunha10b}.
Fig. \ref{fig:duststars} shows ratios of \mdust/\mstar\ plotted against  \mstar\
(in the left panel) and against sSFR (in the right).
There is very little correlation between \mdust/\mstar\ and \mstar, but
\mdust/\mstar\ and sSFR are strongly correlated with 
dust-to-stellar mass ratios increasing with sSFR.
A similar trend was found by \citet[][]{dacunha10b}
for a sample of star-forming galaxies
selected from the Sloan Digital Sky Survey (SDSS), with a requirement 
of 60 and 100\,\micron\ detections from the {\it Infrared Astronomical Satellite}.
The three curves in the right panel Fig. \ref{fig:duststars}
correspond to the 
MS of star formation given by \citet[][and shown in Fig. \ref{fig:ms}]{karim11}, 
but incorporating the regression of \mdust\ with SFR found by 
\citet{dacunha10b}\footnote{\mdust\,(\msun)\,=\,(1.28$\times$2.02)$\times10^7$ [SFR\,(\sfr)$^{1.11}$]
where we have multiplied the \citet{dacunha10b} normalization
by a factor of 2.02 to correct their (lower) dust masses to the \grasil\ scale
because of differences in the assumed dust emissivity.
}.
The vertical trend of the $z\sim2$ (dot-dashed) curve results from the fall-off
of the sSFR at high stellar masses at that redshift.

Interestingly, the correlation between \mdust\ and SFR
(and \mdust/\mstar\ and sSFR)
found by \citet{dacunha10b} overestimates dust mass
relative to stellar mass 
at high sSFR \citep[see also][]{dacunha10a};
this is probably not surprising since the correlation was calibrated in the Local Universe
where such extremes are rare.
As seen also in Fig. \ref{fig:ms}, Fig. \ref{fig:duststars} shows that
GRBHs tend toward high sSFR, and occupy the same region
in parameter space as local ULIRGs \citep{dacunha10a}, 
$z\sim1$ ULIRGs \citep{lofaro13}, and 
$z\sim2$ SMGs \citep{michalowski10}.
In all galaxy populations shown here, at high sSFR $\ga 10^{-9}$\,yr$^{-1}$
the amount of dust compared with stellar content is lower 
than would be expected from local trends in less extreme galaxies.
This is consistent with the results of \citet{hjorth14} who predict
the existence of a ``maximum attainable dust mass'' which causes
a turn-over of the relation at high sSFR.

Figure \ref{fig:dustsfr} shows another permutation of the three variables
\mstar, \mdust, and SFR, namely the ratio, \mdust/SFR, plotted against sSFR.
It can be seen that there is a strong correlation in the sense that
\mdust/SFR decreases with increasing sSFR.
As in Fig. \ref{fig:duststars}, the three curves correspond to the \citet{karim11} MS
together with the \mdust-SFR correlation by \citet{dacunha10b}.
The deviation of this trend with respect to the data 
at high sSFR is even
greater than with the ratio of dust mass to stellar mass and sSFR (see Fig. \ref{fig:duststars}).
At high sSFRs, dust mass relative to SFR is much lower
than the expected, almost constant, trend.
Here, as in previous figures, the GRBHs have properties similar to dusty
massive galaxies (e.g., ULIRGs and SMGs) with high sSFRs. 

The spread in the dust/stars mass ratios and sSFRs in the GRBHs  
is due mostly to redshift.
In Figs. \ref{fig:duststars} and \ref{fig:dustsfr}, the two (IR-detected)
galaxies with the
highest 
dust mass relative to their SFR or sSFR are GRB\,020819B and 090417B, both
relatively local galaxies with $z\sim0.4$ and $z\sim0.3$, respectively.
These hosts have properties very similar to the $z\sim0$ galaxies we have
included in the comparison.
In contrast, the other (IR-detected) GRBHs at similar sSFR 
are at $z\ga1$, and have more than 10 times lower \mdust/\mstar\ and \mdust/SFR.
Indeed, the properties of the GRBHs with $z\ga1$ lie within the range of 
ULIRGs and SMGs; 
more measurements are needed to establish whether a subset of GRBHs 
truly resemble these IR-luminous galaxies as would be suggested by these results.

There are several possible interpretations of the mutual trends of \mdust, \mstar,
and SFR.
\mdust/SFR was interpreted by \citet{dacunha10b} as a proxy for the dust-to-gas mass
ratio of a galaxy because of the gas-SFR scaling relations 
\citep[Schmidt-Kennicutt, e.g.,][]{kennicutt98}: 
\mdust/SFR\,$\propto\,$\mdust/\mgas.
Thus, the inverse correlation between \mdust/SFR and sSFR would imply that 
dust-to-gas ratios are larger in galaxies with low sSFR because the reservoir of
gas available for star formation has been exhausted.
A high sSFR would mean that dust-to-gas ratios are lower because of larger
gas reservoirs, but the dust content would still be high, given the large \mdust/\mstar.
On the other hand, \citet{magdis12} and \citet{sargent13} favor another interpretation,
namely that \mdust/SFR is inversely proportional (together with a metallicity dependence)
to the star-formation efficiency (SFE), defined as the ratio of SFR and gas mass:
\mdust/SFR\,$\propto$\,(Z/Z$_\odot$)\,SFE$^{-1}$\,$\propto$\,(Z/Z$_\odot$)\,\mgas/SFR.
This is the inverse of the dependence on \mgas\ with respect to that hypothesized by \citet{dacunha10b}.
\Citet{magdis12} and \citet{sargent13}
argue that for galaxies on the MS of star formation,
SFEs are relatively constant, and lower than for starbursting systems with high sSFR.
This would explain the negative correlation between \mdust/SFR and sSFR,
and its variation with redshift.

\begin{figure*}[!ht]
\centerline{
\includegraphics[angle=0,height=0.48\linewidth,bb=18 160 592 650]{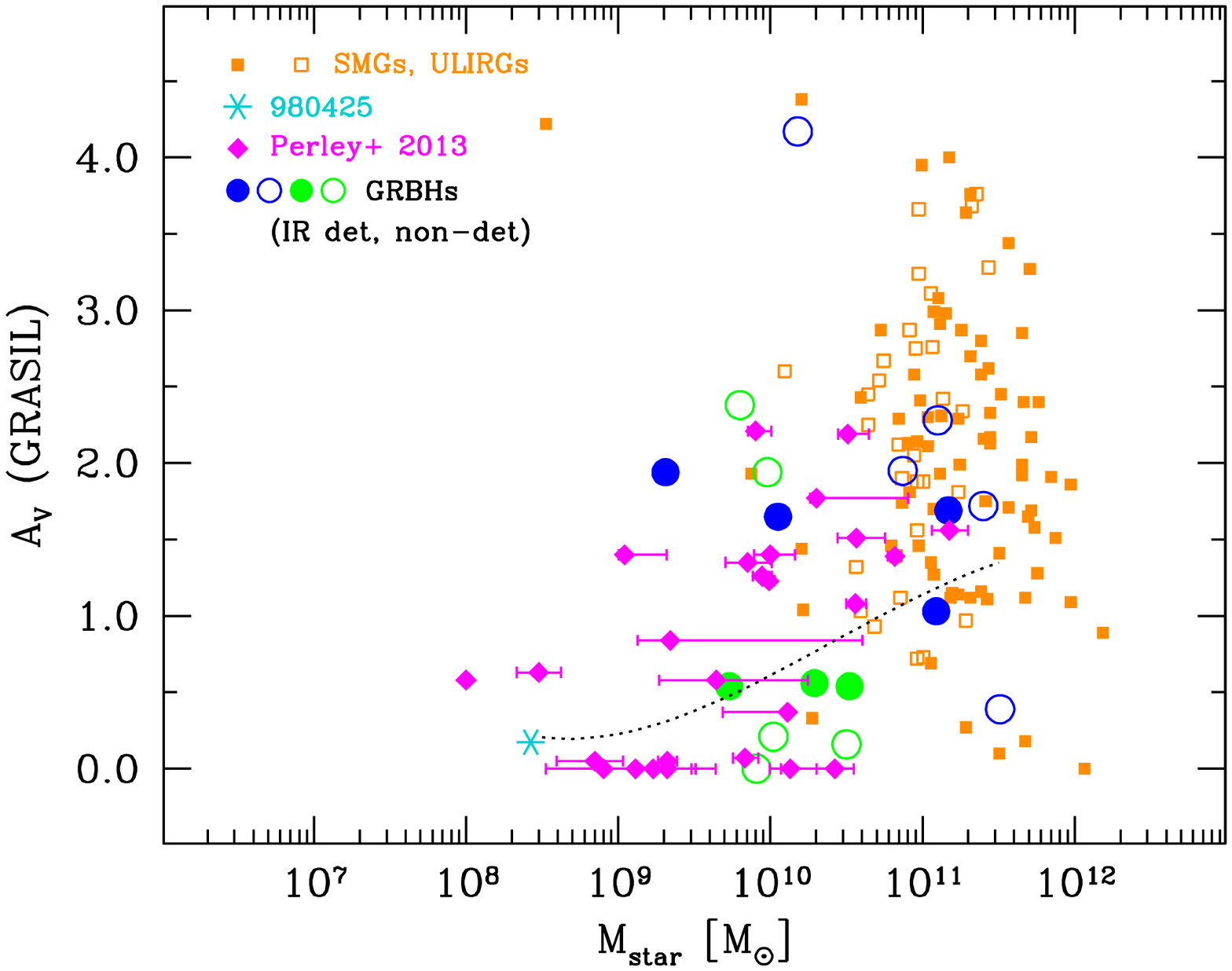}
\hspace{-1.9cm}
\includegraphics[angle=0,height=0.48\linewidth,bb=18 160 592 650]{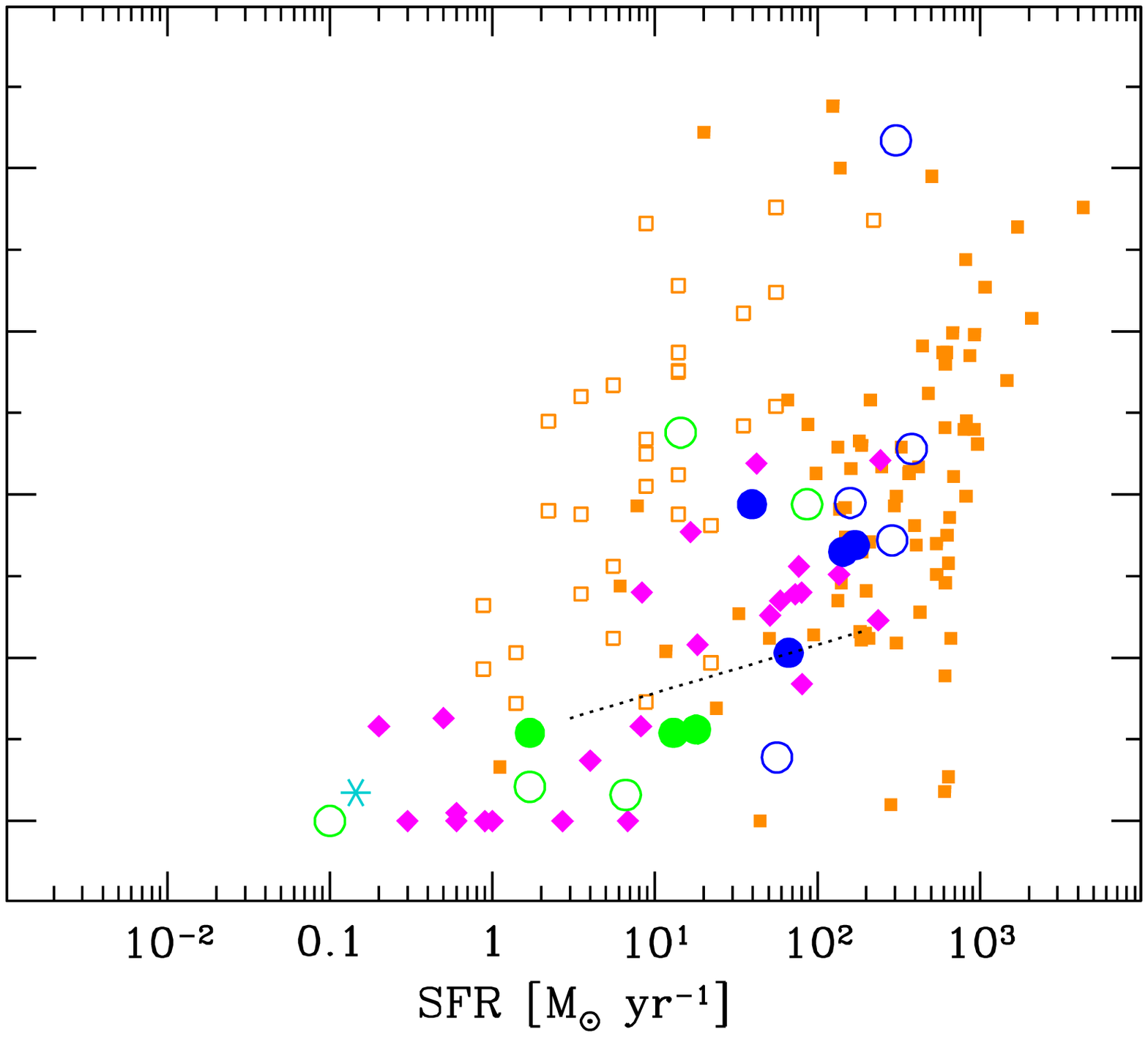}
}
\caption{Extinction \av\ (mag) from \grasil\ fits \citep[except for those from][]{perley13}
plotted against \mstar\ (left panel) and SFR (right).
As in Fig. \ref{fig:ms},
the GRBHs observed by \hers\ are indicated by 
filled (for IR detections) and open (IR non-detections) circles,
and the other galaxy samples shown are also coded as in Fig. \ref{fig:ms}.
The curves show the trends 
for galaxies in the Local Universe \citep[left panel,][]{garn10a}, and 
for $z\sim0.8$ \citep[right,][]{garn10b}.
\label{fig:extinction}
}
\end{figure*}

We propose a third (related) interpretation based on the idea that in the Rayleigh-Jeans
regime, dust mass depends on the 
monochromatic IR luminosity, \lnu, and inversely on the mean dust temperature, \tdust.
Because SFR is linear with \lir\ \citep[assuming an optically thick star-formation
episode, e.g.,][]{kennicutt98}, SFR would scale like \tdust$^{4+\beta}$
since \lir\ results from integrating \lnu\ over frequency $\nu$ \citep[e.g.,][]{hayward11}.
Thus, \mdust/SFR would be inversely proportional to \tdust$^{5+\beta}$.
\tdust\ is expected to vary as some power of the mean radiation field intensity
$\langle U \rangle$ \citep[\tdust$\propto U^{1/(4+\beta)}$, e.g.,][]{hirashita04},
meaning that SFR/\mdust$\propto$\tdust\,$\langle U \rangle$.
This would be consistent with the finding by \citet{magdis12} that \lir/\mdust\ in 
$z\sim2$ IR galaxies increases with $\langle U \rangle$ in the 
\citet{draine07} dust models used to fit their SEDs.
Mean \tdust\ would be higher for galaxies with high sSFR, and could
be a way to distinguish starbursts from MS galaxies as proposed by \citet{hayward12}
and \citet{magnelli13}.
Of the objects in our GRBH sample, the two closest hosts, GRB\,020819B and GRB\,090417B,
are apparently MS galaxies, since their properties are consistent with the
comparison local galaxy populations.
Neither of these objects has any stellar mass produced in a recent burst of star
formation (see Fig. \ref{fig:burst}).
On the other hand, the GRBHs at $z\ga1$ resemble some of the local ULIRGs and $z\sim2$ SMGs;
thus they may be starburst galaxies, at least judging from their position in 
the right panel of Fig. \ref{fig:duststars} and in Fig. \ref{fig:dustsfr}.
Indeed, as suggested by 
\citet{priddey06} and 
\citet{michalowski08,michalowski10}, GRBs may be a way to select
warm SMGs, which before \hers\ were difficult to identify.

\subsection{Extinction from SED fitting}
\label{sec:extinction}

Extinction in galaxies is known to correlate both with stellar mass and with SFR
\citep{garn10a,garn10b},
although since these latter quantities are related through the star-formation MS,
the correlations are probably not independent.
We show extinction \av\ as estimated by the best-fit \grasil\ models 
plotted against  stellar mass in Fig. \ref{fig:extinction} (left panel) and SFR (right)
for the GRBHs and other samples with available data for extinction from SED fitting.
The curve in the left panel is the trend found by \citet{garn10a} for a sample of $z\sim0$
SDSS galaxies using the Balmer decrement to estimate \av, and in the right panel, 
the trend of \av\ with SFR found by \citet{garn10b} for galaxies between $z\sim0$ and $z\sim0.8$.
Both curves are shown over the ranges in parameters for which they were defined.
In Fig. \ref{fig:extinction},
we have converted the empirical \av\ curves for \ha\ to \av\ using the
extinction curve by \citet{cardelli89}.
Because extinction of emission lines tends to be higher than that
of the continuum, we have also applied the recommended
correction factor of 0.44 to both curves
\citep{calzetti00,garn10b}.

Fig. \ref{fig:extinction} shows that for the GRBHs (and the SMGs and ULIRGs),
there is a large scatter in \av\ for a given \mstar\ or SFR, but in general 
\grasil\ \av's for the GRBHs are consistent with those predicted by both curves.
The \grasil\ models give an average optical thickness at 0.55\,\micron, defined
as the (natural logarithm) of the ratio of the dust-free flux and the observed flux. 
We convert this to an optical extinction in magnitudes \av\ by multiplying by 1.086.
Unlike the \grasil\ \av,
the curves shown in Fig. \ref{fig:extinction} are calculated with the Balmer decrement 
which implicitly assumes a screen geometry,
with a dust screen absorbing the stellar radiation between it and the observer.
\grasil\ values of \av, taking into account the dust
and stellar spatial distributions, do not differ in a systematic way from
the values of \av\ inferred by assuming
a dust screen (as is usually assumed in optical-near-IR SED fitting).

\citet{perley13} compared the \av\ in the hosts to the extinction in the GRB afterglow,
and found that in most GRBHs the galaxy-wide extinction is consistent to within
a factor of 2 to the afterglow \av.
This would imply that the dust is distributed in a fairly homogeneous way, and
that local effects are not usually important \citep[although for a counter example see][]{greiner13}.
The Perley et al. targets were selected on the basis of the afterglow extinction
requiring \av$\ga$1\,mag, making their sample a reasonable representation
of dust-obscured GRBs.
In terms of dust properties, it should be very similar 
to our sample which is dominated by the hosts of dark GRBs (14 of 17).
Indeed, relative to our GRBHs,
there is no noticeable difference in the \av's of
the hosts in \citet{perley13}, and 
the GRBHs studied by \citet{perley13} appear to follow expected
trends of \av\ with both \mstar\ and SFR.
As with other properties studied here, the GRBH population (or at least the
dark and/or dusty subset) appears to be normal in terms of dust extinction.

\begin{figure*}[!ht]
\centerline{
\includegraphics[angle=0,height=0.48\linewidth,bb=18 160 592 650]{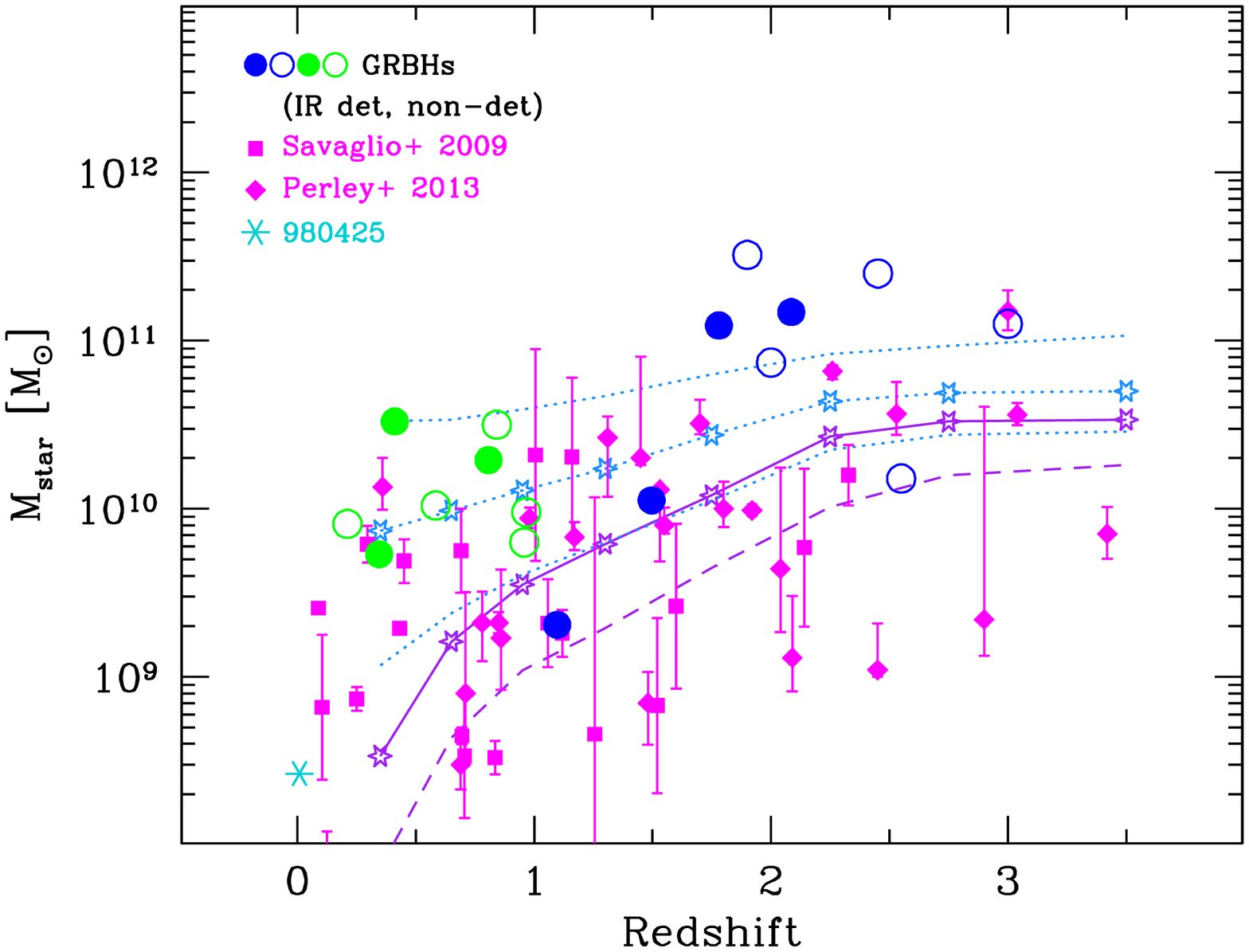}
\hspace{-1.9cm}
\includegraphics[angle=0,height=0.48\linewidth,bb=18 160 592 650]{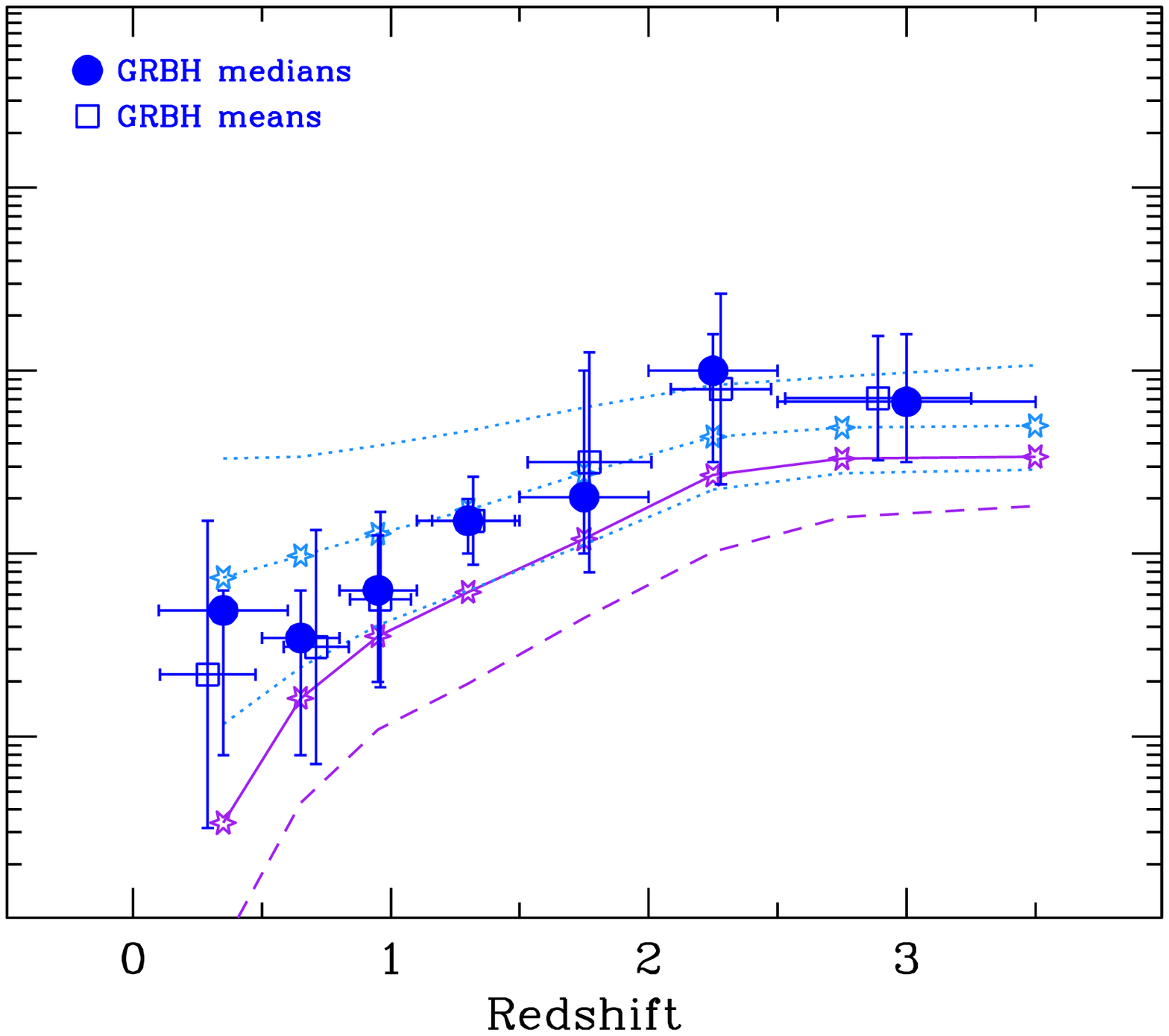}
}
\caption{Stellar mass \mstar\ plotted against redshift.
Individual GRBHs are shown in the left panel, and medians and means 
in the right.
The GRBHs shown in the left panel and observed by \hers\ are indicated by 
filled (for IR detections) and open (IR non-detections) circles,
and the other GRBH samples are coded as in Fig. \ref{fig:ms}.
The curves show the medians of the star-forming population as a function of $z$
(see text).
The solid line connecting open (purple) stars gives the median \mstar\
as a function of $z$, and
the  dotted line connecting the open (light blue) stars corresponds to the
SFR-weighted medians of \mstar\ as described in the text;
the dotted lines below and above these weighted medians show the 
upper and lower quartiles, respectively.
The lowest dashed (purple) line indicates the mass limits of the survey.
In the right panel, the GRBH medians shown as filled circles
are calculated in the same redshift bins as the UltraVISTA comparison \citep{ilbert13}.
The vertical error bars correspond to the 
upper and lower quartiles
of the GRBH distributions, and the horizontal error bars to the width of the
redshift bins. 
The open squares give the GRBH means within each redshift bin, and the error
bars the standard deviation. 
For the GRBH statistics, we considered only the hosts with \mstar\ above the 
\citet{ilbert13} survey lower-mass limit \mlow. 
\label{fig:mass_z}
}
\end{figure*}

\section{GRBHs in context}
\label{sec:context}

In the previous sections, we have analyzed the properties of the GRBHs
in our sample \citep[and when possible also in][]{savaglio09,perley13}, 
and found that their stellar masses, dust masses, SFR, and dust extinction
generally conform to what is known about other star-forming galaxy populations, 
both locally and up to $z\ga3$.

Our \hers\ sample is dominated by hosts of dark GRBs,
and the sample studied by \citet{perley13} was defined on the basis
of high dust extinction (\av$\ga$1\,mag) in the GRB afterglow.
In contrast, the GRBHs analyzed by \citet{savaglio09} come almost exclusively
from optically-bright GRBs.
By combining these three samples (e.g., Fig. \ref{fig:ms}), 
we have a total of 66 GRB hosts (having
eliminated the relatively large overlap among the samples).
To our knowledge, this is the largest long-GRBH sample with reliable stellar masses
and SFRs compiled to date, with redshifts ranging from $z\sim0$ to $z\ga3$ and a median
redshift $z_{\rm med}$\,=\,1.1 (see Sect. \ref{sec:sample}).
These 66 GRBHs consist of:
32 hosts of dark GRBs\footnote{As mentioned in Sect. \ref{sec:sample},
we define a dark GRB as one with optical-to-X-ray spectral index
$\beta_{\rm ox}<$0.5.} (23 of which are hosts of high-\av\ GRB afterglows), 
and 34 
hosts of optically-bright GRBs. 
The resulting fraction of dark GRBs in our combined sample is 48\%,
slightly higher than the upper limit of the possible fraction of dark bursts ($\la$40\%)
reported by \citet{greiner11}.
We conclude that our sample comprises a collection of GRBHs that is, if anything, biased
towards dark and dusty bursts. 
Since there has been a dearth of such GRBs in most samples studied up to now,
it will provide an ideal test for investigating trends of stellar mass and SFR with redshift.

%
\begin{center}
\begin{table*}
      \caption[]{Combined GRBH \mstar, sSFR vs. redshift statistics} 
\label{tab:grbhs_again}
{ 
\begin{tabular}{ccccrrrrrr}
\hline
\multicolumn{1}{c}{Central} & 
\multicolumn{1}{c}{Total} & 
\multicolumn{1}{c}{Log} & 
\multicolumn{1}{c}{Number} & 
\multicolumn{3}{c}{Log(\mstar)} &  
\multicolumn{3}{c}{Log(sSFR)} \\ 
\multicolumn{1}{c}{$z$} & 
\multicolumn{1}{c}{Number$^a$} & 
\multicolumn{1}{c}{\mlow$^b$} & 
\multicolumn{1}{c}{\mstar$\geq$\mlow} & 
\multicolumn{1}{c}{Median} & 
\multicolumn{1}{c}{25$^{\rm th}$ \%} & 
\multicolumn{1}{c}{75$^{\rm th}$ \%} & 
\multicolumn{1}{c}{Median} & 
\multicolumn{1}{c}{25$^{\rm th}$ \%} & 
\multicolumn{1}{c}{75$^{\rm th}$ \%} \\ 
&&&& \multicolumn{1}{c}{(\msun)} & 
\multicolumn{1}{c}{(\msun)} & 
\multicolumn{1}{c}{(\msun)} & 
\multicolumn{1}{c}{(yr$^{-1}$)} & 
\multicolumn{1}{c}{(yr$^{-1}$)} & 
\multicolumn{1}{c}{(yr$^{-1}$)} \\ 
\multicolumn{1}{c}{(1)} &
\multicolumn{1}{c}{(2)} &
\multicolumn{1}{c}{(3)} &
\multicolumn{1}{c}{(4)} &
\multicolumn{1}{c}{(5)} &
\multicolumn{1}{c}{(6)} &
\multicolumn{1}{c}{(7)} &
\multicolumn{1}{c}{(8)} &
\multicolumn{1}{c}{(9)} &
\multicolumn{1}{c}{(10)} \\
\hline
\\
0.35 & 11    &   7.86 & 11  &   9.69 & 8.9 &  9.8  &   $-$9.29 & $-$8.8 & $-$9.5  \\
0.65 & 8     &   8.64 &  6  &   9.54 & 8.9  & 9.8  &   $-$9.11 & $-$8.8 & $-$9.3  \\
0.95 & 10    &   9.04 &  9  &   9.80 & 9.3  & 10.1 &   $-$8.88 & $-$8.5 & $-$9.3  \\
1.30 & 8     &   9.29 &  6  &  10.18 & 10.0 & 10.3 &   $-$9.08 & $-$7.9 & $-$9.9 \\
1.75 & 10    &   9.65 &  8  &  10.31 & 10.0 & 11.0 &   $-$8.98 & $-$8.3 & $-$9.3  \\
2.25 & 8     &  10.01 &  4  &  11.00 & 10.5 & 11.2 &   $-$8.95 & $-$8.8 & $-$9.1  \\
3.00 & 7     &  10.24 &  4  &  10.83 & 10.6 & 11.2 &   $-$8.90 & $-$8.4 & $-$9.3  \\
\\
\hline
\end{tabular}
}
\vspace{0.5\baselineskip}
\begin{description}
\item
[$^{\mathrm{a}}$] 
These numbers do not add up to the total of 66 GRBHs in our combined sample
because three have $z\leq$0.1 (GRB\,060505, GRB\,060218, GRB\,980425)
and have not been included in the statistics.
\item
[$^{\mathrm{b}}$] 
These are the lower limits in the stellar masses
of the survey by \citet{ilbert13}.
\item
[Col. (1)]=\,Central redshift of the bin;
\item
[Col. (2)]=\,Total number of GRBHs within the redshift bin;
\item
[Col. (3)]=\,The lower mass limit \mlow\ of the UltraVISTA survey at this redshift \citep{ilbert13};
\item
[Col. (4)]=\,Total number of GRBHs in this bin with \mstar$\geq$\mlow;
\item
[Cols. (5-7)]=\,Median and percentiles of the GRBH \mstar\ within this redshift bin;
\item
[Cols. (8-10)]=\,Median and percentiles of the GRBH sSFR within this redshift bin.
\end{description}
\end{table*}
\end{center}

\subsection{GRBH stellar masses as a function of redshift}
\label{sec:grbhmass}

Here we examine whether GRBs could be unbiased tracers of 
cosmic co-moving SFR density \rhostar\ by
comparing the stellar masses and sSFRs of our combined GRBH sample with the statistics
of star-forming galaxies found in
recent large-scale deep multi-wavelength surveys.
This will be important to establish whether GRBs
trace SFR to high redshift in an unbiased way, or whether GRBHs are in
some way not representative of typical star-forming galaxy populations.

One of the main arguments against using GRBs to trace \rhostar\
has been that GRBHs tend to be less massive than representative galaxy populations. 
Especially at $z\la1.5$, the host population has been previously
found to favor low stellar masses, and when nebular metallicities
can be measured, also low metal abundance \citep{savaglio09,levesque10a}.
Now, with more data available for hosts of dark and dust-extinguished GRBs, 
we can reassess the question of stellar masses and SFRs in GRBHs.

Our approach is to compare the distributions in redshift of \mstar\ and sSFR
of our GRBH sample with the statistics
of 220\,000 galaxies selected from the deep ($K_s<$24\,AB mag) wide UltraVISTA 
survey \citep{mccracken12,ilbert13}.
The unprecedented depth and coverage of this survey obviates the need to
compensate for cosmic variance, and enables a robust comparison with
the GRBH population.
We make no corrections to the statistics of the GRBH sample, implicitly
assuming that all possible hosts have been identified, that their
redshifts have been determined, and that all GRBs that have exploded have been 
localized, and their host detected.
Clearly none of these assumptions are correct \citep[e.g.,][]{fynbo09,kruhler12,salvaterra12,hjorth12}, but
there is no straightforward method for adjusting any of the statistics of 
our compiled GRBH sample.

As a comparison quantitative benchmark of stellar mass as a function 
of redshift, we have adopted
the double Schechter \citep{schechter76} functions given by \citet{ilbert13}
for the star-forming galaxies in the UltraVISTA survey;
these fits are based on $\sim$135\,000 star-forming galaxies selected
with a two-color criterion (restframe NUV$-r^{+}$, $r^{+}-J$)
to separate them from quiescent galaxies.
At all redshifts, star-forming galaxies are $\ga$6 times more common in number
than quiescent ones, and at redshifts $z\la1.5$, they tend to be a factor of 
$\sim$2 less massive.
At each redshift, to calculate the median stellar mass
we have integrated the double Schechter functions of \citet{ilbert13}
from the lower mass limit dictated by the depth of their survey \mlow\ to $10^{13}$\,\msun.
Because we want to test the hypothesis that GRBs are tracing star formation,
the stellar mass integrands have been weighted by SFR 
as determined from the analytical functions of SFR as a function of redshift 
by \citet{karim11}.
Once we have the integrals as a function of $z$, it is straightforward 
(either numerically or analytically) to calculate the median stellar mass, 
the median stellar mass weighted by SFR, and their 25$^{\rm th}$ and 75$^{\rm th}$
percentile dispersions.

\begin{figure*}[!ht]
\centerline{
\includegraphics[angle=0,height=0.48\linewidth,bb=18 160 592 650]{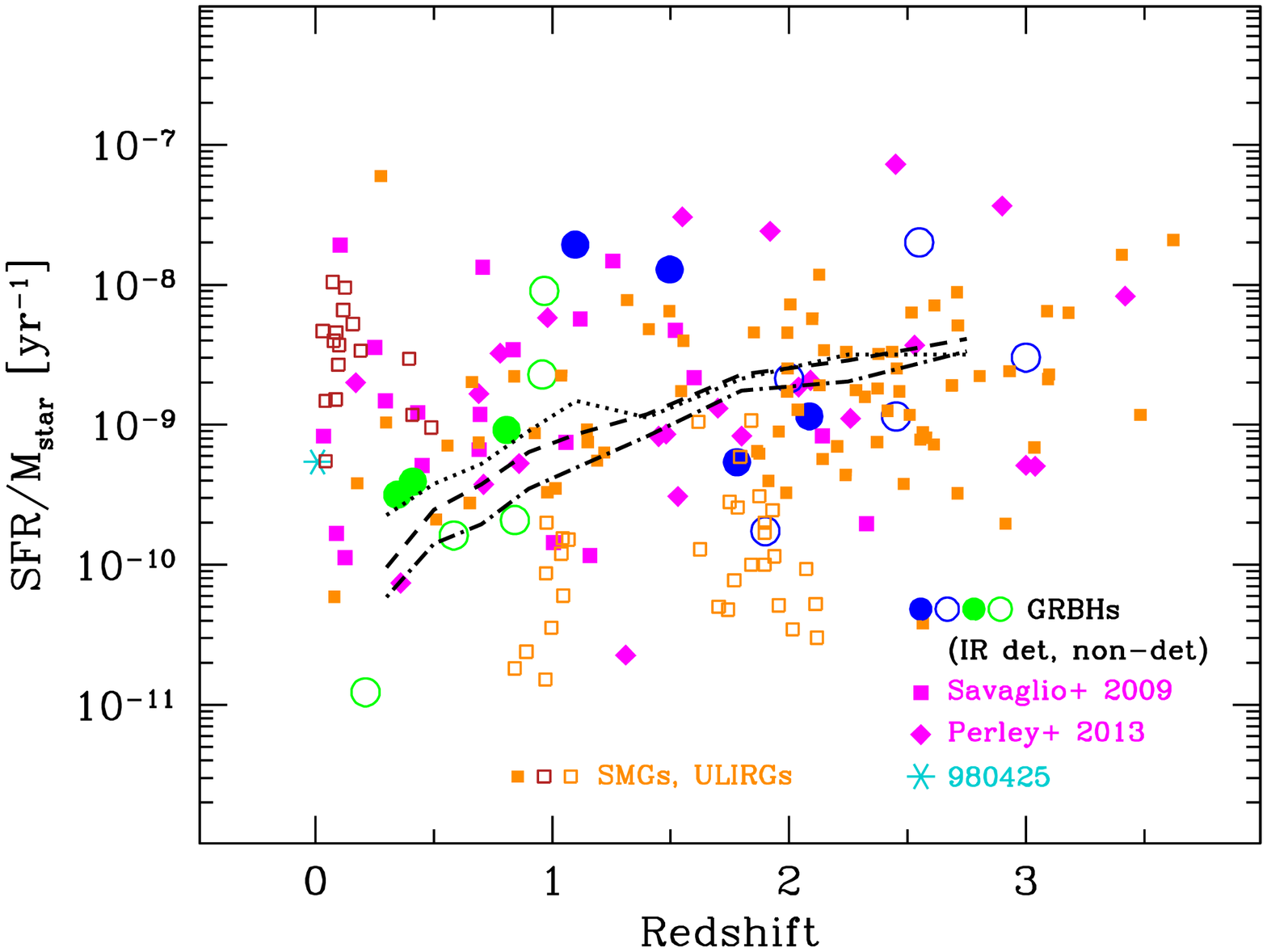}
\hspace{-1.9cm}
\includegraphics[angle=0,height=0.48\linewidth,bb=18 160 592 650]{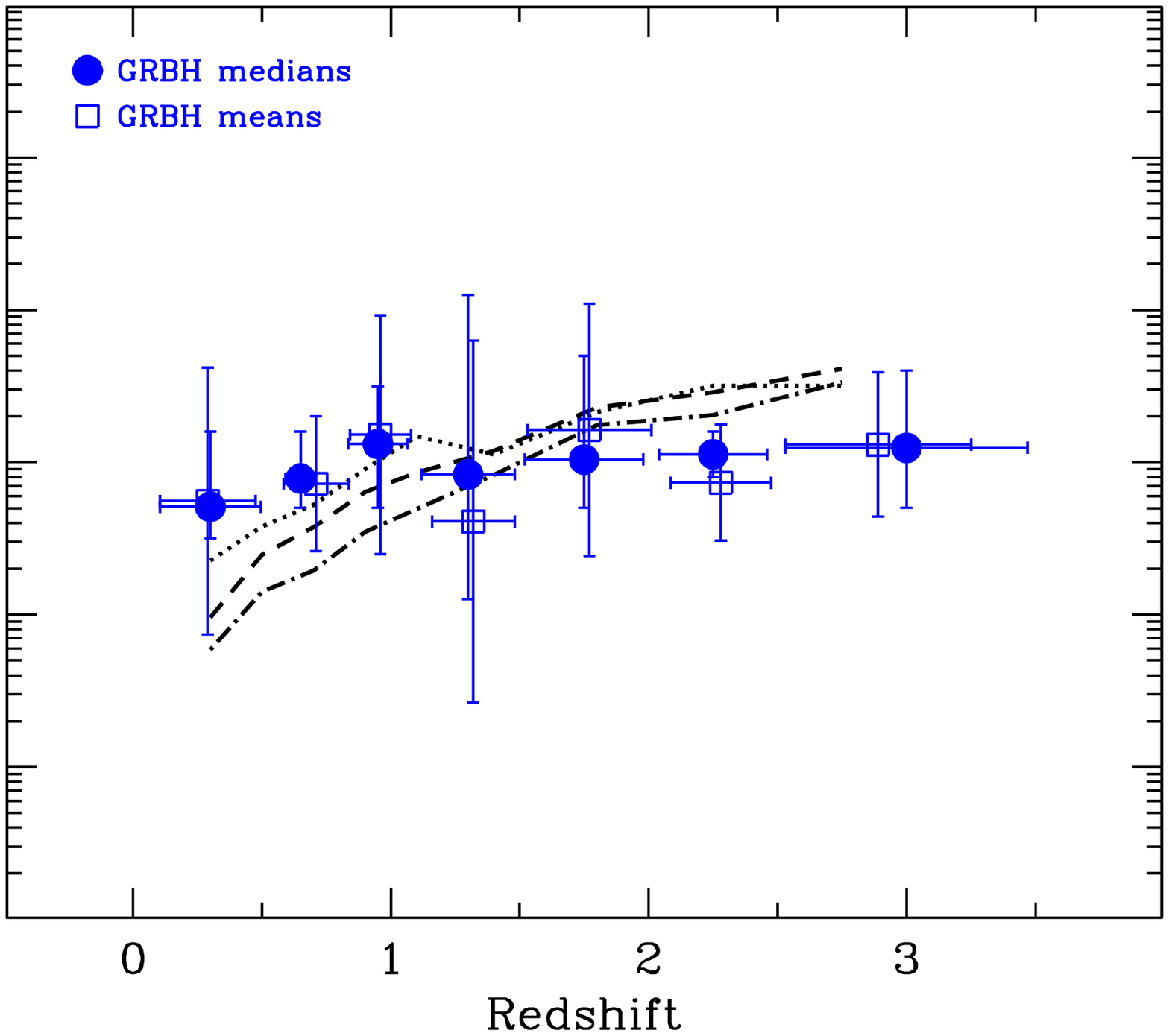}
}
\caption{sSFR SFR/\mstar\ plotted against redshift.
Individual GRBHs, SMGs, and ULIRgs are shown in the left panel, and medians and
means of the GRBH sample in the right.
As in Fig. \ref{fig:ms},
in the left panel the GRBHs observed by \hers\ are indicated by 
filled (for IR detections) and open (IR non-detections) circles,
and the other GRBH samples shown are also coded as in Fig. \ref{fig:ms}.
The curves show the results from \citet{karim11}, 
and correspond to three values of log(\mstar)\,=\,9.6 (dotted line), 
10.4 (dashed), and $>$11\,\msun\ (dot-dashed).
The GRBH medians in the right panel, shown as filled circles
for the lowest-mass bins at a given redshift, and 
and filled squares for the highest-mass bins as described in the text.
The vertical error bars correspond to the 
upper and lower quartiles
of the GRBH distributions, and the horizontal error bars to the width of the
redshift bins (for the GRBH data).
Open squares show the means of the GRBH distributions within each redshift bin, and 
the error bars correspond to the standard deviation.
As in Fig. \ref{fig:mass_z},
for the GRBH statistics,
we considered only the hosts with \mstar\ above the 
\citet{ilbert13} survey \mlow\ limit. 
\label{fig:ssfr_z}
}
\end{figure*}

Figure \ref{fig:mass_z} shows the results of these calculations graphically; 
the left panel shows the individual GRBHs, and the medians and means of the host
distribution are shown on the right.
Table \ref{tab:grbhs_again} reports the statistics of the combined GRBH sample
shown in the figure;
the GRBH medians, means, and 
dispersions
are calculated without considering the hosts with \mstar\ below \mlow\
of the survey [see Col. (2) of Table \ref{tab:grbhs_again}].
The curves in both panels of Fig. \ref{fig:mass_z} correspond to median \mstar\
(solid line) and the median SFR-weighted \mstar\ (dotted line) of the UltraVISTA survey, as
inferred from integrating the double Schechter functions given by \citet{ilbert13}.
The medians of SFR-weighted \mstar\ fall above the non-weighted medians because 
more massive galaxies have higher SFR, thus skewing the median masses
to higher values \citep[e.g.,][]{karim11}.
The left panel of Fig. \ref{fig:mass_z} shows clearly that beyond $z\sim0.5$,
GRBs can identify galaxies of such low mass that they fall well below the
limits of even the deep UltraVISTA survey (as shown by the dashed line).

At all redshifts, 
the GRBH median stellar masses fall close to or slightly below the expected median of the 
SFR-weighted star-forming galaxies at those redshifts.
One possible exception is the $z\sim2$ bin, where the GRBH median and mean \mstar\ both lie
{\it above} the SFR-weighted median of the UltraVISTA distribution.
Taking the results at face value, and within the mass limits of UltraVISTA,
the distributions (medians and dispersions) of the GRBH \mstar\ all fall within the range of 
50\% of the UltraVISTA distribution of normal star-forming galaxies. 
Thus we would conclude that there is no strong evidence that the host population is 
not representative of typical star-forming galaxies; on the contrary GRBHs are
apparently similar
to the more general star-forming galaxy population at least for $z\la3$.

This result contrasts with \citet{perley13} who found that at $z\la1.5$,
there is a ``highly significant aversion" to massive GRBHs, namely a preference for
low-mass systems that exceeds expectations for a purely SFR-selected sample.
The most likely reason for the difference between our conclusions and theirs 
is the different comparison samples and the comparison methodology. 
\citet{perley13} use the narrow but deep survey ($K_s<$23\,Vega mag, 24.8\,AB mag) 
by \citet[][ MOIRCS Deep Survey, MODS]{kajisawa09}, slightly deeper than the UltraVISTA
survey but with $\sim$25 times fewer galaxies in the $z=0.5-1.0$ redshift bins
because of the narrow field-of-view.
Although the depth of the two surveys is comparable, the statistical accuracy is 
much greater with UltraVISTA.
Moreover,
the MODS survey does not distinguish between quiescent and star-forming galaxies,
making it impossible to compare GRBHs with only the star-forming population.
As mentioned above, at redshifts $z\la1.5$ 
star-forming galaxies tend to be less massive by a factor of 2 or more
than quiescent populations.
Our results show that the GRBHs are as massive or even more massive than the 
(non-SFR-weighted) star-forming galaxy populations at all redshifts; this is a robust result
(see Fig. \ref{fig:mass_z}) and shows the importance of weighting \mstar\ with SFR. 
Another possible reason for the contrast
is that many of the least massive hosts fall below
the mass limits of the UltraVISTA survey \citep[and the MODS survey][]{kajisawa09}, 
and we thus did not consider them in the GRBH statistics.
Although \citet{perley13} checked this and concluded that completeness was not a problem for
the MODS comparison, combined with other factors this could also contribute to the
difference between our conclusions.
A final possible reason is the double-episode SFH incorporated by our \grasil\ fitting, 
as opposed to the single star-formation episode used by \citet{perley13};
as mentioned in Sect. \ref{sec:comparison}, this is known to give larger stellar masses
\citep{michalowski12a}.

We do find, however, a possible trend of host masses being less massive 
then the UltraVISTA SFR-weighted
\mstar\ median up to $z\la1$, but the dispersions are large.
Besides \citet{perley13}, many previous studies
\citep[e.g.,][]{lefloch03,fynbo03,boissier13} 
also found that GRBHs are a biased representation of star-forming galaxies 
because they tend to be less massive and more metal-poor for $z\la1.5$
\citep[although see][]{michalowski12b}.
However, such conclusions are probably sample-dependent.
We find that the host masses as a function of redshift
depend on the percentage of dark or dusty bursts within a given redshift bin.
At all redshifts, the hosts of dark bursts are more massive than
the median GRBH \mstar\ at that redshift\footnote{The host of GRB\,070306
is an exception to this, having \mstar\ roughly equal to the median
of the distribution.}.

Even though our sample is relatively large, the numbers are still 
sparse when distributed in redshift space.
Indeed, the largest discrepancy relative to the star-forming weighted median \mstar\
is the $z=0.65$ bin; 
it contains only two hosts of dark GRBs 
which are also the most massive in that bin.
Because of the small number (4) hosts of non-dark GRBs in the
same redshift bin,
one more dark GRB host at a similar mass ($\sim10^{10}$\,\msun)
could have raised the median 
to $\sim$dex(9.8)\,\msun\ (the same median mass as the $z=0.95$ bin), 

\subsection{SFR and sSFR of GRBHs as a function of redshift}
\label{sec:grbhssfr}

We now examine trends of sSFR with redshift for our sample and compare them
with the properties of the COSMOS sample analyzed by \citet{karim11}.
Figure \ref{fig:ssfr_z} shows sSFR of the GRBHs and $z\sim1-2$ ULIRGs and SMGs
in the left panel plotted against redshift, and in the right panel
the GRBH medians and mean sSFR.
The curves in both figures correspond to the trends of sSFR with $z$ for three
specific values of \mstar\ [dex(9.6)\,\msun, dex(10.4)\,\msun, $>$dex(11)\,\msun]
as found by \citet{karim11}. 
Because the strongest trend of sSFR is with redshift rather than with \mstar\
\citep[e.g.,][]{karim11},
to calculate the medians for the GRBHs we have binned only in redshift,
rather than again sub-dividing into mass bins; this helps conserve statistical
significance given our relatively small sample size.
To be consistent with the \mstar\ statistics, we considered only the hosts with
\mstar\ larger than the UltraVISTA lower-mass limit.
Columns 8-10 of Table \ref{tab:grbhs_again} give the median sSFR as a function
of $z$ for our host sample.

For $z\la1$, the median sSFRs of the hosts are always higher than the curves
given by \citet{karim11}.
This tendency to high sSFRs is almost certainly related to the tendency toward
lower \mstar\ seen in Fig. \ref{fig:mass_z}.
Beyond $z\sim1$, the median GRBH sSFR is similar to that expected values for massive
galaxies, again consistently with the trends for \mstar\ shown in Fig. \ref{fig:mass_z}.
We conclude that at low redshift, $z\la1$, GRBHs tend to have slightly lower \mstar,
and slightly higher sSFRs than the star-forming galaxy populations at those redshifts.
However, the difference is dominated by large scatter, and disappears 
entirely beyond $z\ga1$.

\subsection{Reconciling our results}
\label{sec:reconciling}

In this paper, we have focused on dust mass, stellar mass and sSFRs, 
and have concluded that, although there is large dispersion,
the GRBH population appears to have similar characteristics as the more
general star-forming galaxy population for $z\la3$.
However, there are numerous studies based on other aspects of GRBs
and their host galaxies that reach a different conclusion.
Here we attempt to summarize these and formulate a view consistent with
all the evidence.

The supposed inability of GRBHs to trace cosmic SFR is most
pronounced at low redshift, $z\la1$.
In this redshift range, the GRB property that has been most under scrutiny 
is metallicity, namely the finding that $z\la1$ GRBHs are more metal poor 
than general star-forming galaxy populations 
\citep[e.g.,][]{savaglio09,han10,levesque10a}.
For a given \mstar,
metallicities of nearby GRBHs are found to fall below 
the SFR-weighted mass-metallicity relation \citep{kocevski11,graham13}.
On the other hand,
\citet{mannucci11} find that GRBHs differ from the mass-metallicity
relation because of their high SFRs, not because their O/H is anomalously
low for their \mstar.
Indeed, many GRBHs have high sSFR, and would be considered starbursts since they
lie well above the MS of star-formation (see e.g., Fig. \ref{fig:ms}).
Because of the so-called
Fundamental Metallicity Relation or Fundamental Plane of \mstar, O/H, and SFR 
\citep[e.g.,][]{mannucci10,hunt12}, 
at a given \mstar, GRBHs are thus expected to be more metal poor 
than galaxies along the MS. 
The relatively low abundances and high sSFRs of GRBHs essentially
trace a ``starburst sequence'' of
the mass-metallicity relation as shown by \citet{mannucci11}.

In any case, if there is any metallicity bias at $z\la1$, 
it appears to set in at relatively high metallicities
\citep[$Z\la$50\%\,\zsun,][]{hao13,graham13},
and is quantitatively small \citep{wolf07}. 
There is possibly some evidence that such a bias could change with redshift 
up to $z\la$ 1 \citep{boissier13}.
Theoretical models show only a moderate preference for low metallicity
\citep{trenti13}, although even this could arise from a bias because of
using only redshifts measured from optical afterglows \citep{wanderman10}.

In fact, most of the above studies were based on relatively small samples
of GRBHs with clearly detected optical afterglows
\citep[e.g.,][]{han10,kocevski11,graham13}.
The problem with analyzing metallicity in galaxies at $z\la1$ is that 
optical spectra are needed, which requires afterglows
or host galaxies that are sufficiently bright to observe spectroscopically.
Hence, such analyses tend to exclude, by design and by necessity, dark hosts, 
thus possibly compromising the general applicability of the conclusions about
trends with metallicity. 

Another aspect of $z\la1$ GRBs that could indicate that they are not unbiased
tracers of star formation is their environments and the kinds of galaxies
that they reside in.
The environments of nearby long GRBs have been found to differ from those of
core-collapse supernovae (CCSNe), leading to the conclusion that GRBs
are associated with only the most massive stars and tend to reside in
fainter and more irregular galaxies than the hosts of CCSNe
\citep{fruchter06}.
However, this result is not confirmed by \citet{kelly08} and \citet{leloudas10}
who found
that type Ic SNe, those associated with long GRBs, are more likely to erupt 
in the brightest regions of their hosts, similarly to GRBs themselves.
Nevertheless, the relations of stellar mass with
galaxy morphology, SFR, and metallicity make 
it difficult to establish whether low-redshift GRBs 
tend to occur in low-mass galaxies because they are metal poor
or because such galaxies dominate star-forming populations at these
redshifts.
Given that GRBs select galaxies with a SFR weighting scheme,
the predominance of metal-poor low-mass galaxies at $z\la1$ could simply be 
a consequence of the high fractions of low-mass (\mstar$\leq 10^{10}$\msun), 
blue, high sSFR (``high activity'') 
systems which are known to dominate the star-forming galaxy population 
for $0.5\la z \la 1$ \citep{ilbert10}.

At redshifts $z\ga1$, 
the main evidence pointing toward a potential inability of GRBH to trace cosmic SFR
density, \rhostar, is based on observed cumulative redshift distributions.
In contrast to the smooth decline in \rhostar\ for $z\ga3$ 
inferred from UV continuum surveys \citep[e.g.,][]{hopkins06,bouwens11,ellis13},
GRB rates need \rhostar\ to remain flat to $z\ga8$
\citep[e.g.,][]{kistler09,butler10,robertson12,jakobsson12}.
This could be because the bulk of star formation at high redshifts
occurs in galaxies that are below the detection limit of even the deepest UV surveys 
\citep[e.g.,][]{tanvir12,trenti12,trenti13};
such a faint population would play an important role in 
cosmic reionization \citep{finkelstein12}.
Moreover, when a series of biases are taken into account when comparing
the rate of GRBs and \rhostar, there is apparently no discrepancy between
the normal star-forming galaxy population and the hosts traced by GRBs
\citep{elliott12}.
This debate is far from being resolved, but wider and deeper
galaxy surveys together with further analysis of unbiased GRB samples will 
provide more constraints.

Finally, there is the ISM content of GRBHs which in part we have examined
in this paper through \hers\ observations.
Because star formation is associated with molecular gas,
and because of the high sSFRs in GRBHs, GRB host galaxies would be expected to 
have a fairly robust molecular hydrogen \htwo\ content.
However, tracing \htwo\ through GRB absorption-line or damped \lya\ systems (DLAs)
has been quite difficult.
Molecular gas at $z\ga2$ has been detected so far in only two and possibly a third 
GRB-DLA \citep{prochaska09,fynbo06,kruhler13}.
One of these (GRB\,080607) is metal-rich, and one is metal-poor (GRB\,120815A),
so there is no clear trend even with these small numbers.
Because sightlines with low dust depletion (and low metallicity) favor
the detection of DLAs, selection effects have significant impact 
on the study of \htwo\ in GRB-DLAs \citep{kruhler13}.
Larger samples of afterglows observed with powerful instruments such
as the VLT/X-Shooter and ALMA will undoubtedly help better understand 
the molecular content in GRBHs.

\section{Summary and conclusions}
\label{sec:conclusions}

For the first time, we are able to place constraints on dust emission in a significant
sample of the hosts of long GRBs.
We have observed with \hers\ 17 GRBHs from $z\sim0.2$ to $z\sim3$, and detected
dust emission in 7.
The probability of IR detection in the dark subset of our sample is 43\%, compared to
$\la$20\% from previous attempts to detect IR emission which targeted hosts of
optically-bright GRBs \citep[e.g.,][]{tanvir04,lefloch06}.
Fitting the multiwavelength SEDs with \grasil\ gives dust and stellar masses, SFRs, and dust
extinction, and enables a comparison of these properties in GRBHs with other star-forming
galaxy populations at similar redshift.
This comparison shows that GRBHs at $z\ga$0.5 tend to be galaxies with high sSFRs
and high dust-to-stellar mass ratios, but are in any case similar in terms
of \mstar, SFR, and \av\ to other populations.
The \mdust/\mstar\ ratios and sSFR of GRBHs are similar to SMGs and ULIRGs at similar redshifts.
The trends of \mdust/SFR suggest that
GRBs may select warm SMG-like objects, which before \hers\ were
difficult to identify.

To investigate whether GRBs 
can be used to trace co-moving SFR in the universe,
we have incorporated additional GRBHs in our analysis and compared them with 
trends of redshift for \mstar\ and sSFR as given by the recent UltraVISTA survey of 220\,000
galaxies ($\sim$135\,000 star-forming).
The results show that the stellar masses of GRBHs in our combined sample
lie within the range of 50\% of the UltraVISTA star-forming galaxy population from $z\sim$0.3 to $z\sim3$,
although there is a large dispersion.
The sSFRs of the GRBHs tend to be high, but
are also within the range of values expected for star-forming galaxies
at similar redshifts.
Thus we conclude that GRBs select galaxies that are representative
of the more general population; 
hence they should trace the cosmic SFRD in an unbiased way.

Our sample is dominated by hosts of dark GRBs, and dark hosts have 
\mstar\ above the GRBH median in all redshift bins.
Because dark and dusty bursts tend to be found in more mature, metal-rich
galaxies, it 
is possible that this sample composition is driving our conclusion
that GRBH are unbiased tracers of star formation.
Better statistics are needed, together with careful SED fitting,
to better address the properties of the hosts of dark GRBs,
their impact on the GRBH population as a whole, 
and whether the hosts of GRBs are generally representative of high-$z$ galaxies.

\begin{acknowledgements}
We warmly thank Ruben Salvaterra for enlightening discussions,
and Iv\'an Oteo for giving us the data for the LBG $z=1,2$ samples.
We are grateful to the anonymous referee for insightful and timely comments that
improved the clarity of the manuscript.
MJM and is a postdoctoral researcher of the FWO-Vlaanderen (Belgium), and
acknowledges the support of the Science and Technology Facilities Council.
AR acknowledges support by the Th\"uringer Landessternwarte Tautenburg,
and SB, VDE, LKH, EP, and AR are grateful to support from PRIN-INAF 2012/13.
SS acknowledges support from the Bundesministerium f\"ur
 Wirtschaft and Technologie through DLR (Deutsches Zentrum f\"ur 
Luft- und Raumfahrt e.V.) FKZ 50 OR 1211, and
PS acknowledges support through the Sofja Kovalevskaja Award from
the Alexander von Humboldt Foundation of Germany.
Part of the funding for GROND (both hardware as well as personnel) was 
generously granted from the Leibniz-Prize to Prof. G. Hasinger 
(DFG grant HA 1850/28-1). 
Use was made of the NASA/IPAC Extragalactic Database (NED),
and the public website for GRB host galaxy data ({\tt http://www.grbhosts.org/})
maintained by S.~Savaglio and collaborators.
\end{acknowledgements}

\newpage

\begin{appendix}

\section{Photometry tables of the GRBHs}
\label{app:tables}

See published version.
%

\end{appendix}

\end{document}